\documentstyle[12pt]{article}
  
\newlength{\extralineskip}
\newcommand{\beq}{\begin{equation}}
\newcommand{\eeq}{\end{equation}}
\newcommand{\bd}{\begin{displaymath}}
\newcommand{\ed}{\end{displaymath}}

\def\e{~{\rm e}}
\def\bea{\begin{eqnarray}}
\def\eea{\end{eqnarray}}

\def\ba{\beq\new\begin{array}{c}}
\def\ea{\end{array}\eeq}

\def\inbar{\,\vrule height1.5ex width.4pt depth0pt}
\def\IC{\relax\hbox{$\inbar\kern-.3em{\rm C}$}}
\def\IR{\relax{\rm I\kern-.18em R}}

\def\IN{\relax{\rm I\kern-.15em N}}

\makeatletter
\newdimen\normalarrayskip              
\newdimen\minarrayskip                 
\normalarrayskip\baselineskip
\minarrayskip\jot
\newif\ifold             \oldtrue            \def\new{\oldfalse}
\def\arraymode{\ifold\relax\else\displaystyle\fi} 
\def\@arrayskip{\ifold\baselineskip\z@\lineskip\z@
     \else
     \baselineskip\minarrayskip\lineskip2\minarrayskip\fi}
\def\@arrayclassz{\ifcase \@lastchclass \@acolampacol \or
\@ampacol \or \or \or \@addamp \or
   \@acolampacol \or \@firstampfalse \@acol \fi
\edef\@preamble{\@preamble
  \ifcase \@chnum
     \hfil$\relax\arraymode\@sharp$\hfil
     \or $\relax\arraymode\@sharp$\hfil
     \or \hfil$\relax\arraymode\@sharp$\fi}}
\def\@array[#1]#2{\setbox\@arstrutbox=\hbox{\vrule
     height\arraystretch \ht\strutbox
     depth\arraystretch \dp\strutbox
     width\z@}\@mkpream{#2}\edef\@preamble{\halign \noexpand\@halignto
\bgroup \tabskip\z@ \@arstrut \@preamble \tabskip\z@ \cr}%
\let\@startpbox\@@startpbox \let\@endpbox\@@endpbox
  \if #1t\vtop \else \if#1b\vbox \else \vcenter \fi\fi
  \bgroup \let\par\relax
  \let\@sharp##\let\protect\relax
  \@arrayskip\@preamble}

\begin{document}
\thispagestyle{empty}

\begin{center}
{\huge \bf Loop Correlators and Theta States in 2D Yang-Mills
Theory}\footnote{ This work is supported in part by the Natural
Sciences and Engineering Research Council of Canada, the Istituto
Nazionale di Fisica Nucleare and NATO CRG 930954. The work of L.P. is
supported on part by a UBC University Graduate Fellowship.}\\
\vskip 0.3 truein
{\bf G. Grignani$^{(a)}$, L. Paniak$^{(b)}$, G. W. Semenoff$^{(b)}$ and P.
Sodano$^{(a)}$}
\vskip 0.3truein
$(a)$ {\it Dipartimento di Fisica and Sezione
I.N.F.N., Universit\`a di Perugia, Via A. Pascoli I-06123 Perugia,
Italy}\\
\vskip 1.truecm
$(b)$ {\it Department of Physics and Astronomy\\University of British 
Columbia\\
6224 Agricultural Road\\Vancouver, British Columbia, Canada V6T 1Z1}\\
\vskip 0.5truein
DFUPG-97-158; UBC-GS-8-97
\vskip 0.5truein
{\bf Abstract}
\vskip 0.3truein
\end{center}
Explicit computations of the partition function and correlation
functions of Wilson and Polyakov loop operators in theta-sectors of
two dimensional Yang-Mills theory on the line cylinder and torus are
presented.  Several observations about the correspondence of two
dimensional Yang-Mills theory with unitary matrix quantum mechanics
are presented.  The incorporation of the theta-angle which
characterizes the states of two dimensional adjoint QCD is discussed.

\newpage
\setcounter{page}1

\section{Introduction}
Two dimensional (2D) Yang-Mills theory is an interesting example of a
topological field theory~\cite{wit,blau0,blau}.  It has been used
extensively as a toy model where features that it shares with higher
dimensional gauge theories can be studied in a simplified
context~\cite{cmr,ab}.  As a quantum field theory, it has no
propagating degrees of freedom.  It nevertheless exhibits a confining
potential for heavy quarks at the tree level.  It is also an example
of a gauge theory where (in a planar geometry) the Migdal-Makeenko
equations for Wilson loops can be solved explicitly.  Correlators of
Wilson loops can be computed in closed form ~\cite{mig,rus} and the
area law associated with confinement physics can be demonstrated
explicitly. Moreover, it has been demonstrated that the strong
coupling, large $N$ limit of 2D Yang-Mills theory on the sphere can be
rewritten as a random surface model~\cite{gt1,gt2}. This supports the
long-standing conjecture that the infrared limit of a confining gauge
theory is related to string theory.  Its dynamics are also known to be
intimately related with D=1 unitary matrix models which are related to
two dimensional string theories \cite{mp1,mp2,dadda}.
 
As a topological field theory, 2D Yang-Mills theory gives a field
theoretical example of localization formulae \cite{wit,blau0}.  It is
a quantum field theory whose partition function is given exactly by
evaluating the Euclidean functional integral in the saddle point
approximation.  The technique which is used to compute the functional
integral is the so-called diagonalization method where gauge
invariance is used to diagonalize the matrix-valued degrees of freedom
to yield a simple, solvable model for the eigenvalues.  The spectrum
which is obtained this way differs from the conventional one.  This
difference is related to the existence of inequivalent quantizations
of the gauge theory, depending on the order in which the constraints
are imposed and quantization is done~\cite{hh1,hh2,hh3,hh4}, 
and was noted by
Witten in his analysis of 2D Yang-Mills theory~\cite{wit}.
The diagonalization procedure has also been used to compute the
partition function and correlators of Polyakov loop operators in ref.
~\cite{gss,wipf}.  

In this Paper we shall summarize some of our observations about 2D
Yang-Mills theory.  We shall begin by making a few remarks about the
correspondence between Yang-Mills theory and unitary matrix quantum
mechanics.  We shall make extensive use of one such well-known
correspondence ~\cite{r3,mp1,r1,r2} where the unitary matrices are the Wilson
loops.  We shall also formulate another one where the dynamical
variables are unitary matrices related to Polyakov loops.  These
matrices determine the gauge group holonomy that the wave-functions of
heavy quarks accumulate in the quantized Yang-Mills theory.  The
latter correspondence is related to recent work where it was shown
that the question of whether or not a gauge theory with adjoint matter
and at finite temperature and density exists in a confined or
de-confined phase could be related to the question of symmetry
breaking in a certain nonlinear sigma model
\cite{gss,gsst1,gps}.  In that model, the dynamical variable is a
unitary matrix whose trace coincides with the Polyakov loop operator.
The Polyakov loop operator is an order parameter for breaking the
discrete symmetry and its expectation value characterizes confinement
in finite temperature adjoint gauge theories \cite{pol,sus}.  In
general the action of the non-linear sigma model in question is a
complicated effective action coming from integrating gauge and matter
fields from the functional integral \cite{sy1,sy2}.  However, it was argued
in ref.  ~\cite{gss} that the effective action has a simple form in 2D
Yang-Mills theory and the resulting model could be solved exactly
using the diagonalization method ~\cite{blau2}.  This idea was
generalized to Yang-Mills theory with external sources in ~\cite{stz}
where it was used to show that there is a de-confining phase
transition at infinite $N$ limit of a hot and dense gas of heavy
quarks.

In this Paper, we shall complete the program begun in ref. ~\cite{gss}
of presenting a full computation of the correlators of loop operators
in 2D Yang-Mills theory using the diagonalization method.  We are
particularly interested in the sectors of the theory with non-trivial
theta vacua.  We will present an explicit computation of the two-point
correlators of Wilson and Polyakov loop operators in a theta-sector of
2D Yang-Mills theory at finite temperature.  We shall consider both
the case where the space is a circle, so that the spacetime is a
torus, or the space is a line with fixed boundary conditions, so that
the spacetime is a cylinder.  We shall compute the partition function,
find the behavior of loop correlators and deduce the quark-antiquark
potential for both of these cases.  We show that, in the limit of
infinite volume, our result reproduces the already known
facts~\cite{wit2,psz1,psz2} that, in a sector of the theory with
non-zero theta angle, fundamental representation quarks have a
repulsive interaction and adjoint quarks have a screened, short-ranged
interaction.  Our results for the correlators of Polyakov loop
operators are different from those found in previous computations of
these quantities using the diagonalization technique ~\cite{gss,wipf}.

Our principal result is the demonstration that the diagonalization
technique can be modified in such a way that it obtains the
conventional spectrum of 2D Yang-Mills theory on a circle.
Furthermore, we demonstrate that the two-point correlators which are
obtained by this method agree in various limits with those obtained by
the group theoretical character expansion technique.  The distinct
advantage of the diagonalization method is that it can be used to
obtain explicit formulae for correlators of higher winding loops.

\subsection{Theta states}

It was first argued by Witten \cite{wit2} that 2D Yang-Mills theory,
or 2D QCD with adjoint matter, has theta states.  Since the Yang-Mills
fields transform in the adjoint representation of the gauge group $G$,
the true gauge group is the factor group of $G$ by its center, $C$,
$G/C$.  Generally, the resulting group is not simply connected.  For
example, if $G$ itself is simply connected ($\Pi_1(G)=0$).  
which we shall henceforth
assume, then
\begin{equation}
\Pi_1(G/C)=\Pi_0(C)=C
\end{equation}
This has non-trivial consequences for any gauge theory that is defined
on a spacetime which itself has non-trivial fundamental group.  For
the moment, let us assume that the space is a circle, $S^1$.\footnote{
We could as well use the real line $R^1$.  In the case of the real
line, the gauge group is taken as the set of those gauge
transformations which go to the identity at infinity.} The group of
time-independent gauge transformations, ${\cal G}$ , which is formed
from the smooth mappings of the circle, $S_1$, to $G/C$ has the
property
\begin{equation}
\Pi_0({\cal G})= \Pi_1( G/C)=C
\end{equation}
Since the Hamiltonian is gauge invariant, and the quantum states are
invariant under gauge transformations in the component of the gauge
group ${\cal G}$ which contains the identity, the quantum states must
carry an irreducible unitary representation of the Abelian group $C$.
All such representations are one-dimensional.  Each distinct
representation is called a theta-state (this terminology derives from
the representations of elements of U(1) by phases $e^{i\theta}$).  

All gauge invariant operators have vanishing matrix elements between
states which carry different representations of $C$.  Thus, there is a
superselection rule which allows one to choose one particular
representation to characterize all of the quantum states of the
theory.

For example, if $G=SU(N)$ then its center is $C=Z_N$, the cyclic 
group of order $N$ and 
\begin{equation}
\Pi_0({\cal G})=\Pi_1\left( SU(N)/Z_N\right)=Z_N
\end{equation} 
There are N irreducible unitary representations of the center, $Z_N$,
which have elements
$$1,e^{i\theta},e^{2i\theta},\ldots,e^{i(N-1)\theta}$$ where the $N$
allowed values of $\theta$ are $$ \theta = 0,2\pi/N,...,2\pi
k/N,...,2\pi(N-1)/N $$

The partition function of Yang-Mills theory is
\begin{equation}
Z=\int dA_\mu \exp\left( -\int_M d^2x\frac{1}{2e^2}{\rm Tr}(F_{\mu\nu}^2)
\right)
\end{equation}
Let us consider the case where the Euclidean space-time is a Riemann
surface, $M$.  The gauge fields belong to topological sectors.  In a
generic topological sector, the connection one-form $A$ is not
globally defined, but is a smooth 1-form only in coordinate patches.
Consider a triangulation of the space $M$ by a set of simply connected
subspaces $T^I$ labeled by $I$ ($M=\cup_I T^I$) and let the
connection in patch $I$ be $A^I$.  At the boundary between patches $I$
and $J$, the connections are related by a gauge transformation
$A^I=(A^J)^{g^{IJ}}$.

We can change a gauge field which is in one topological sector to one
in another sector by the following surgical procedure~\cite{wit}:
Consider a particular field configuration $A_I$ in patch $I$ and
transition functions $g_{IJ}$ which relates $A_I$ to the connections
in neighboring patches.  We can change topological sectors by
replacing $A_I$ by its gauge transform $A^g$ and the transition
functions $g_{IJ}$ by $g_{IJ}g$ if we choose $g$ to have the property
that the restriction of $g$ to the boundary of the patch (which is
isomorphic to $S^1$) is an element of ${\cal G}$, the smooth mappings
of $S^1$ to $G/C$, which is in a non-trivial component of $\Pi_0({\cal
G})=C$.  The resulting field configuration will be in a different
topological sector from the original one.  The number of topological
sectors is equal to the number of elements in $C$.  Summing over the
topological sectors in the path integral with complex
unimodular weights given by the corresponding elements of the
irreducible unitary representation of the center, $C$, produces the
partition function in a $\theta$ sector.

The above arguments depend on certain smoothness requirements for the
class of field configurations which contribute to the path integral
and also of the gauge transformations which are allowed in the
Hamiltonian formalism.  However, as independent evidence for the
existence of theta states, they have been given a physical
interpretation similar to the theta-vacuum in the Schwinger-model
~\cite{coleman1,coleman2} as being created by the existence of charges at the
boundaries of the space.  The theta parameter is known to affect the
spectrum of mesons in 2D QCD with heavy quarks ~\cite{wit2,psz1}.
Arguments of stability of such systems have also been shown to be
consistent with the topological classification of theta
states~\cite{wit2,psz2}.

\subsection{Overview and Summary of Results:}

In Section 2, we shall review the quantization of Yang-Mills theory in
1+1-dimensions when the space is a circle.  We pay particular
attention to the existence of large gauge transformations and the
associated appearance of $\theta$ vacua.  Then, in Section 3, we
derive a number of different representations of the heat kernel of 2D
Yang-Mills theory.  We also present an explicit calculation of the
heat kernel using the diagonalization technique.  Section 4 contains a
computation of loop correlators on the cylinder and torus.  We also
obtain results for SU(2) and compare them with a similar computation
using character expansion techniques and find that they agree.

We summarize our results as follows:
\begin{itemize}
\item{}We obtain the partition functions and the
correlators of two loop operators for 2D QCD on both the cylinder and
the torus in a $\theta$-state when $G=SU(N)$ given in formulae
(\ref{kercyl}), (\ref{ztorus}), (\ref{corcyl}) and (\ref{cortoro}).
This is done using the representation of the heat kernel in
(\ref{heatkernel}) below.  In the process, we show that the trace of
(\ref{heatkernel}) leads to the character representation of the
partition function (\ref{charrep}).
  
\end{itemize}
It is well known\cite{r3} that the physical states of 2D Yang-Mills
theory on a circle ($x\in[0,L]$) are class functions of the Wilson
loop $$
\psi_{\rm phys}(U)=\psi_{\rm phys}(vUv^{\dagger})
$$ where $vv^{\dagger}=1$ and $$ U(\tau)~=~{\cal P}\exp\left(i\int_0^L
dx A_1(x,\tau)\right) $$ At finite temperature where the Euclidean
time is periodic ($t\in[0,\tau=1/T]$), the Polyakov loop operator is
also of interest, $$ g(x)~=~{\cal P}\exp\left( i\int_0^{1/T} dt
A_0(x,t)\right) $$ The heat kernel of 2D Yang-Mills theory, defined by
$$
\langle \psi_2\left| e^{-H\tau}\right|\psi_1\rangle~=~
\int[dU_1][dU_2]\psi_2^*(U_2)\psi_1(U_1)K[\tau;U_2,U_1]
$$ so that $$ K[\tau;U_2,U_1]=\langle U_2\left| e^{-H\tau} \right|
U_1\rangle $$ is defined as a symmetric function of the eigenvalues of
$U_2$ and $U_1$ separately.  It can be written in three ways: It has
the well known decomposition in terms of characters $\chi_R(U)={\rm
Tr}_RU$ of $U$ in irreducible representations $R$, $$
K[\tau;U_2,U_1]~=~\sum_R \chi_R^*(U_2)\chi_R(U_1)e^{-e^2L\tau C_2(R)}
$$ where $C_2(R)$ is the second Casimir invariant of $R$.  In
addition, we shall find that it can be presented
\begin{itemize}
\item{}as a D=1 unitary matrix model with a background gauge field,
$$ K[\tau;U_2,U_1]=\int[dg(x)]\exp\left(-\frac{1}{2e^2\tau}
\int_0^Ldx{\rm Tr}\left|\nabla g+iA_2g-igA_1\right|^2\right)
$$ where $g(0)=g(L)$ and $$ U_{1}~=~{\cal P}e^{i\int_0^L dx A_1(x,0)}
~~,~~ U_{2}~=~{\cal P}e^{i\int_0^L dx A_1(x,\tau)} $$
\item{}as a D=1 gauge invariant matrix model
\begin{eqnarray}
K[\tau;U_2,U_1]&=& \int [dU(t)]dA(t)\exp\left(-\frac{1}{2e^2L}
\int_0^\tau dt{\rm Tr}\left|\dot U-i[A,U]\right|^2\right)\cdot  
\nonumber\\ && \cdot\,
\delta_{\rm cl}(U(\tau),U_2)\delta_{\rm cl}(U(0),U_1)J(U(\tau))J(U(0))
\label{heatkernel}
\end{eqnarray}
where $J(U)$ is the Vandermonde determinant and $\delta_{\rm cl}(U,V)$
is the conjugation invariant delta function which equates the spectra
of $U$ and $V$.
\item{}The partition function of Yang-Mills theory in a $\theta$-state 
on the 2-sphere with area $L\tau$ can be written as the D=1 unitary
matrix model $$ Z[{\rm sphere}~;\theta]~=~\sum_z{\cal
Z}(z,\theta)\int[dg_z]\exp
\left(-\frac{1}{2e^2\tau}\int_0^L
{\rm Tr}\left| \nabla g_z\right|^2\right)
$$
where $g_z(0)=zg_z(L)$, $z$ is an element of $C$, the center of $G$, 
and 
$Z(z,\theta)$ is the element of the $\theta$-representation of $C$ 
corresponding to $z$.
\item{}The partition function of Yang-Mills theory in a $\theta$-state 
on the 2-torus with area $L\tau$ can be
written as the gauge invariant D=1 matrix model
$$
Z[{\rm torus}~;\theta]~=~\sum_z{\cal Z}(z,\theta)\int[dg_z][dA]\exp
\left(-\frac{1}{2e^2\tau}\int_0^L
{\rm Tr}\left| \nabla g_z-i[A,g_z]\right|^2\right)
$$

In the equivalent character representation the partition function of
Yang-Mills theory on a Riemann surface of genus $g$ and in a
$\theta$-sector is given by
\begin{equation}
Z[g,\theta]=\sum_R \left( {\rm dim~R}\right)^{2-2g}
\delta(R,R_\theta)\exp\left(-e^2\tau L C_2(R)
\right)
\label{charrep}
\end{equation}
where $R_\theta$ are those representations which have the property
\begin{equation}
\chi_{R_\theta}(zU)={\cal Z}(z,\theta)\chi_{R_\theta}(U)
\end{equation}
\end{itemize}

\section{Hamiltonian formulation of Yang-Mills theory on the circle}
\setcounter{equation}{0}

In this Section we shall review the Hamiltonian formulation of 2D
Yang-Mills theory.  Canonical quantization of the action\footnote{Note
that we use an unconventional normalization of the charge $e^2$.}
\begin{equation}
S~=~\int dxdt~\frac{1}{2e^2}{\rm Tr}\left( \dot A_1-
\nabla A_0\right)^2
\end{equation}
yields dynamical variables which are the spatial component of the
gauge field $A_1(x)\equiv A(x)$ and the electric field $E(x)\equiv
\dot{A}(x)-\nabla A_0(x)$, both of which are matrices in the
fundamental representation of the Lie algebra of the gauge group, and
can be expanded in a canonical set of generators as $$
A(x)=\sum_aT^a
A^a(x)~~,~~E(x)=\sum_a T^aE^a(x) $$
For a compact semi-simple Lie
algebra, $${\rm Tr} T^a T^b= \frac{1}{2} \delta^{ab} ~~,~~ \left[
  T^a,T^b \right]=if^{abc}T^c $$
The Hamiltonian is
\begin{equation}
H= {2e^2}\int_0^L dx {\rm Tr}\left( E^2(x)\right)
\label{ham}
\end{equation}
and the non-vanishing canonical commutation relation is
\begin{equation}
\left[ A^a(x), E^b(y) \right] =i\delta^{ab}\delta(x-y)
\label{com}
\end{equation}
$A_0$ is a Lagrange multiplier which enforces Gauss' law as the
constraint is
\begin{equation}
\nabla E(x)+i\left[ A(x), E(x)\right]~\sim~0
\label{gl}
\end{equation}
and all variables have periodic boundary conditions,
\begin{equation}
E(L)=E(0)~,~A(L)=A(0)
\end{equation}

The gauge transformation is
\begin{equation}
E(x)\rightarrow E^g(x)=g(x) E(x) g^{\dagger}(x)
~~,~~
A(x)\rightarrow A^g(x)=
g(x)\left( A(x)-i\nabla\right)g^{\dagger}(x)
\end{equation}
The periodic boundary conditions for $A(x)$ and $E(x)$ are preserved
by gauge transformations where the matrix $g(x)$ and its first
derivative are periodic up to an element of the center of the gauge
group,
\begin{equation}
g_z(L)=zg_z(0)
\label{gbc}
\end{equation}
where $z\in C$.  When $z=1$, $g_1(x)$ is said to implement a ``small
gauge transformation'' and when $z\neq 1$, $g_z(x)$ is said to
implement a ``large gauge transformation''.

The operator on the right-hand-side of the constraint (\ref{gl})
generates infinitesimal small gauge transformations.  As a result of
the constraint (\ref{gl}), the physical phase space is the set of
equivalence classes of configurations $E(x)$ and $A(x)$ where fields
are in the same equivalence class if they are related by a small gauge
transformation (i.e. one 
with a periodic gauge function and 
with $z=1$ in
(\ref{gbc}) ).  The constraint could be solved at the classical
level by choosing representatives of the equivalence classes.  This
procedure has been discussed in ~\cite{ls1,ls2}.  Alternatively, after
finding the quantum realization of the un-constrained theory with
commutator (\ref{com}) and Hamiltonian (\ref{ham}), the constraint
(\ref{gl}) could be imposed as a physical state condition which
chooses a subspace of the quantum states as `physical states'.  It
implies that physical states of the theory are invariant under all
gauge transformations which can be generated by iterating
infinitesimal transformations, i.e. all small gauge 
transformations.  The coset group of all gauge transformations modulo
periodic ones is isomorphic to the center of the gauge group, $C$.
The physical states must transform under an irreducible unitary
representation of this coset.

In the functional Schr\"odinger picture, where states are
wave-functionals, $\Psi[A]$, of gauge field configurations and the
electric field is realized as
\begin{equation}
E^a(x)~\Psi[A]~=~\frac{1}{i}\frac{\delta}{\delta A^a(x)}~\Psi[A]
~~~,
\end{equation}
the physical state condition,
\begin{equation}
\left( \nabla E(x)+i\left[ A(x),E(x)\right]\right)
~\Psi_{\rm phys.}[A;\theta]~=~0~~~,
\end{equation}
implies that the wave-functionals of physical states transform as
\begin{equation}
\Psi_{\rm phys.}[g_z(A-i\nabla)g_z^{\dagger};\theta]
~=~{\cal Z}(z,\theta)~\Psi_{\rm phys.}[A;\theta]
\label{phys.}
\end{equation}
where $g_z(x)$ has the boundary condition in (\ref{gbc}) and ${\cal Z}
(z,\theta)$ is the representative of the center element $z$ in the
representation labeled by $\theta$.  Since the center of the gauge
group is an Abelian discrete group, all irreducible unitary
representations are one-dimensional and the number of inequivalent
representations is equal to the order of the group.  The solution of
(\ref{phys.}) is
\begin{equation}
\Psi_{\rm phys.}[A;\theta]~=~\psi[U;\theta]
\end{equation}
where $U$ is the unitary matrix
\begin{equation}
U~\equiv~ {\cal P}\exp\left( i\int_0^L dxA(x)\right)
\label{uni}
\end{equation}
with the additional constraint that the wave-function is a class
function of $U$:
\begin{equation}
\psi[U;\theta]~=~\psi[hUh^{\dagger};\theta]
\end{equation}
for any $h$ in the fundamental representation of $G$.  This implies
that the wave-function is a symmetric function of the eigenvalues of
$U$.  On the wavefunctions, we must impose the further requirement
that they lie in a particular theta-sector,
\begin{equation}
\psi[zU;\theta]~=~{\cal Z} (z,\theta)~\psi[U,\theta]
\end{equation}
Let us consider the operator $U_R$ constructed from the path-ordered
phase integral with the gauge field in an arbitrary irreducible
representation $R$ of the Lie algebra,
\begin{equation}
U_R~\equiv~{\cal P}\exp\left( i\int_0^L dx A^a(x)T^a_R\right)
\end{equation}
where $T^a_R$ are the generators in representation $R$.  The electric
field operates on $U_R$ as
\begin{equation}
E^a(x)~U= {\cal P}\exp\left( i\int_x^LdwA^a(w)T^a_R\right)~T^a_R~
 {\cal P}\exp\left( i\int_0^xdw A^a_R(w)\right)
\end{equation}
and the field space Laplacian which is in the kinetic term in the
Hamiltonian as
\begin{equation}
-\left(\frac{\delta}{\delta A^a(x)}\right)^2~U_R=C_2(R)U_R
\end{equation}
where $C_2(R)$ is the second Casimir invariant corresponding to the
representation $R$ which is obtained from the formula
\begin{equation}
C_2(R)\cdot {\cal I}~=~\sum_a T^a_RT^a_R
\end{equation}
We thus find physical eigenstates of the Hamiltonian by forming traces
of $U_R$, i.e., the characters which are defined by
\begin{equation}
\chi_R(U)={\rm Tr}~U_R
\end{equation}
so that
\begin{equation}
H~\chi_R(U)~=~e^2L~C_2(R)~\chi_R(U)
\end{equation}
Thus, the action of the Hamiltonian on the characters is proportional to
that of the group Laplacian, $\Delta(G)$ whose spectrum is the set of
second Casimir invariants of irreducible representations.

Characters are orthonormal with respect to integration over the Haar
measure,
\begin{equation}
\int [dU]~\chi_R(U)~\chi^*_{R'}(U)~=~\delta_{R,R'}
\end{equation}
This inner product can be obtained from the natural inner product,
which is functional integration over $A$, by gauge fixing,
\begin{eqnarray}
\int dA(x)\Psi_{\rm phys.}^*[A]\Psi_{\rm phys.}[A]
={\rm const.}\cdot\int [dU]~\psi^*(U,\theta)\psi(U,\theta)
\label{gf}
\end{eqnarray}

Products of characters are also important as they effectively 
carry out the multiplication of irreducible representations
\beq
\chi_{R_1}(U)\chi_{R_2}(U) = \chi_{R_1 \otimes R_2}(U)
\eeq
If $R_1 \otimes R_2 = \oplus N^{R_3}_{R_1 R_2} R_3$
is a decomposition into
irreducible representations of the product then by linearity of 
characters
\beq
\chi_{R_1}(U)\chi_{R_2}(U) = \oplus N^{R_3}_{R_1 R_2} 
\chi_{R_3}(U)
\eeq
Consequently we have the definition of the fusion numbers, 
\beq
N^{R_3}_{R_1 R_2} = \int[ dU] 
\chi_{R_1}(U)\chi_{R_2}(U)\chi^*_{R_3}(U)
\eeq

Thus the set of all states is identified with the characters of $U$.
The theta-states in a given theta-sector are a subset of these.  For a
certain representation of the center of the gauge group, with elements
${\cal Z}(z,\theta)$ we must chose the characters for those
representations of $G$ which have the property
\begin{equation}
\chi_R(zU)~=~ {\cal Z}(z,\theta)~\chi_R(U)
\end{equation}
The spectrum of the Hamiltonian is then the set of second Casimir
invariants corresponding to the representations $R$ of the group $G$
whose characters have this property.

In the following we shall deal mostly with the special case where the
gauge group is SU(N).  Then the $N^2-1$ generators $T^a$ are $N\times
N$ traceless Hermitean matrices.  The center of SU(N) is the cyclic
group of order $N$, $Z_N$.  The representations which have the
property $\chi_R(e^{2\pi i n/N}U)= {\cal Z}(\theta) \chi_R(U)=
e^{in\theta}\chi_R(U)$ with $\theta=2\pi k/N$ are those
representations whose Young tableaux have $k$ modulo $N$ boxes. If we
consider $U(N)$ representations, this is equivalent to keeping only
those with linear Casimir $C_1(R)=k \mbox{mod}N$.

It is well known that 2D Yang-Mills theory is equivalent to a 
particular
version of unitary matrix quantum mechanics.  This can either 
be seen by fixing a particular gauge as in
\cite{mp1,mp2} or by the following argument which shows that the 
Hilbert space of states, energy levels and degeneracies of the two
theories are identical of we identify the unitary matrices as the
Wilson loop operator in (\ref{uni}).  The Hamiltonian of unitary
matrix quantum mechanics is proportional to the group Laplacian,
\begin{equation}
H_{QM}=e^2L\Delta(G)
\end{equation}
can be derived by canonical quantization of the action
\begin{equation}
S~=~\frac{1}{2e^2L}\int dt~{\rm Tr}\left(\dot U^{\dagger}\dot U\right)
~=~\frac{1}{4e^2L}\int dt~{\rm Tr}\left( iT^aU^{\dagger}(t)
\dot U(t)\right)^2
\end{equation}
The canonical momenta are given by
\begin{equation}
\Pi^a=\frac{1}{2e^2L}{\rm Tr}\left(iT^a U^{\dagger}\dot U\right)
\end{equation}
and the Hamiltonian is
\begin{equation}
H_{QM}~=~e^2L\sum_a (\Pi^a)^2~=~e^2L\Delta(G)
\end{equation}
The canonical momentum operators have the Lie algebra
\begin{equation}
\left[ \Pi^a ,\Pi^b\right]~=~if^{abc}\Pi^c
\end{equation}
and 
\begin{equation}
\left[ \Pi^a,U\right]~=~ T^aU~~,~~\left[ \Pi^a,U^{\dagger}
\right]~=~-U^{\dagger}T^a~~,~~
\end{equation}

The wave-functions of this system are the group elements in unitary
irreducible representations of the group,
\begin{equation}
\Delta(G)~U_R~=~C_2(R)~U_R
\label{laplacian}
\end{equation}
The inner product is given by the integration over the group with the
invariant Haar measure, where the wave-functions have the property
\begin{equation}
\int [dU]~(U_R^{\dagger})_{kl}~(U_R)_{ij}~=~
\frac{1}{{\rm dim}~R}~\delta_{il}\delta_{jk}
\end{equation}
Each component of the unitary matrix in the representation $R$ is a
linearly independent, normalizable wave-function.  The degeneracy of
each eigenstate is equal to the number of linearly independent
components, i.e. $({\rm dim}~R)^2$.

However, to produce the spectrum and degeneracies of 2D Yang-Mills
theory, the wave-functions must be restricted to class functions of
group elements so that, for each representation of $G$. only the
wave-function $\chi_R(U)$ is allowed.  At the operator level, this
restriction is realized by the constraint
\begin{equation}
\sum_a\Pi^a\cdot{\rm Tr}\left( T^aT^b-T^aUT^bU^{\dagger}\right)~\sim~0~~,
\end{equation}
The operator in this constraint generates infinitesimal adjoint
transformations of the group elements.  This constraint can be
enforced by a Lagrange multiplier and the Hamiltonian and constraint
can be obtained by canonical quantization of the action
\begin{equation}
S~=~\frac{1}{2e^2L}\int dt{\rm Tr}\left| \dot U(t)-i\left[
U(t),A(t)\right]\right|^2
\label{mma}
\end{equation}
This action has a gauge invariance under
\begin{equation}
A\rightarrow A^g=gAg^{\dagger}-ig\dot g^{\dagger}
~~,~~
U\rightarrow U^g = gUg^{\dagger}
\end{equation}
Here, the gauge field $A(t)$ is a Lagrange multiplier
enforcing the matrix mechanics analog of Gauss' law. 
The constraint implies that the 
adjoint action of the group generators on the wave-function 
vanishes.   
Wave-functions  which are annihilated by the constraint are class
functions of the group variables,
\begin{equation}
\xi[U]~=~\xi[gUg^{\dagger}]
\end{equation}
Of the $\left({\rm dim}(R)\right)^2$ linearly independent
eigenfunctions $U_R$ of $\Delta(G)$, only one, the character
$\chi_R(U)={\rm Tr}U_R$ has this property.  

In this theory, the expectation value of the Wilson loop operator in a
given quantum state is given by the fusion number,
\begin{equation}
\left< R_1\left| \chi_R(U)\right|R_2\right>~=~
\int [dU]\chi^*_{R_1}(U)\chi_R(U)\chi_{R_2}(U)
~=~N^{R_1}_{R_2 R}
\end{equation}
Similarly, the correlator of a product of Wilson loop operators in
arbitrary representations is given by
\begin{equation}
\left<R\left| \chi_{R_1}(U)\chi_{R_2}(U)\ldots\chi_{R_k}(U)
\right|R'\right>~=~
\int [dU]  \chi^*_R(U)\chi_{R_1}(U)\chi_{R_2}(U)\ldots\chi_{R_k}(U)
\chi_{R'}(U)
\end{equation}
For the group SU(N), explicit formulae for these moments can be
obtained.  Also, the partition function for this theory at finite
temperature $T=\tau^{-1}$ is given the the familiar expression for the
partition function of Yang-Mills theory on the torus,
\begin{equation}
Z[\tau]~=~\sum_R ~\exp\left(-e^2L\tau C_2(R)\right)
\end{equation}
We would expect that this partition function could also be 
written as
a functional integral for the d=1 unitary matrix 
model with action(\ref{mma}).  
We shall show in the following sections that, modulo some
subtleties with boundary conditions, this is indeed the case.

\section{The heat kernel: Yang-Mills theory on the cylinder}
\setcounter{equation}{0}

In this Section, we shall give two equivalent presentations of the
heat kernel for 2D Yang-Mills theory.  The first shows the connection
between 2D Yang-Mills theory and a certain principal chiral model
which generalizes a result in \cite{gss}.  
In this representation, the dynamical variables are the unitary matrices 
which would parallel transport heavy quark wave functions from the initial 
to the final (Euclidean) time. If we consider a partition function, so that 
Euclidean time is periodic, their traces would be Polyakov loop operators.

Alternatively, we shall find it useful to represent the heat kernel 
as a $c=1$ unitary matrix model where the matrices are Wilson loop variables.
We shall show, with
details in subsection \ref{careqpcm}, the equivalence of the latter model
with the well known character expansion of the heat kernel.  

We begin by considering the propagation function in Yang-Mills theory,
\begin{equation}
K[\tau;A_2,A_1]~\equiv~ \left< A_2\left| e^{-H\tau} P \right| A_1 \right>
~~,
\end{equation}
where $\left| A\right>$ is an eigenstate of the gauge field operator
$A^a(x)$, $H$ is the Hamiltonian in (\ref{ham}) and $P$ is a
projection operator onto gauge invariant states.  For the moment, we
consider states which are invariant under periodic gauge
transformations only.  The projection can be implemented by gauge
transforming the field $A_1$ at one side of the trace and integrating
over all gauge transformations,
\begin{equation}
K[\tau;A_2,A_1]~=~\int [dg(x)]\left<A_2\left| e^{-H\tau}\right|A_1^g\right>
\label{hk1}
\end{equation}
(where we normalize the measure so that $\int [dg(x)]=1$ and
$g(0)=g(L)$.)  The integrand in (\ref{hk1}) is the heat kernel which
obeys the following equation,
\begin{equation}
\left( \frac{\partial}{\partial \tau}-e^2\int_0^Ldx\sum_a
\left(\frac{\delta}{\delta A_2^a(x)}\right)^2\right)
\left< A_2\left|e^{-H\tau}\right| A_1\right>=0
\end{equation}
with the boundary condition
\begin{equation}
\lim_{\tau\rightarrow 0}
\left< A_2\left| e^{-H\tau}\right| A_1^g\right>=\prod_x
\delta\left( A_2(x)-A_1^g(x)\right)
\end{equation}
These are solved by\footnote{Here, we have dropped a zero point energy
term for the Hamiltonian.  Also, this equation should be divided by
the normalization factor $(2\pi e^2\tau)^{L\delta(0)}$.  If we use
zeta-function regularization, this normalization has a very simple
form.  Since $$L\delta(0)=\lim_{s\rightarrow 0}(1+2\sum_1^\infty
1/n^s)=0$$ the normalization factor is one.}
\begin{equation}
K[\tau;A_2,A_1]=\int [dg(x)]~\exp\left(-\frac{1}{2e^2\tau}
\int_0^L dx~ {\rm Tr}(A_2-A_1^g)^2\right)
\label{hk2}
\end{equation}
We can re-arrange the action in (\ref{hk2}) to put it in the following form,
\begin{equation}
K[\tau;A_2,A_1]=\int [dg(x)]~\exp\left( -\frac{1}{2e^2\tau}
\int_0^Ldx{\rm Tr}\left|\nabla g+
ig(x)A_1(x)-iA_2(x)g(x)\right|^2\right)
\label{hk3}
\end{equation}
This is the path integral for a 0+1-dimensional principal chiral model
with external gauge fields where we treat the spatial variable $x$ as
Euclidean time.  With this identification, it also coincides with the
partition function unitary matrix quantum mechanics coupled to
external gauge fields.  Since the Haar measure has the properties
$[dg(x)] = [d(u(x)g(x))]= [d(g(x) v^{\dagger}(x))]$ where $u(x)$ and
$v(x)$ are unitary matrices, the heat kernel has the property that it
is invariant, under gauge transformation of $A_1(x)$ and $A_2(x)$
separately,
\begin{equation}
K[\tau;A_2^u,A_1^v]=K[\tau;A_2,A_1]
\end{equation}
This implies that the heat kernel is a class function of each of the
two unitary matrices
\begin{equation}
U_{1}~=~{\cal P}\exp\left(i\int_0^Ldx A_{1}(x)\right) ~~,~~
U_{2}~=~{\cal P}\exp\left(i\int_0^Ldx A_{2}(x)\right)
\label{unitary}
\end{equation}
Consequently we define 
\beq
K[\tau;A_2,A_1]=K\left[\tau ;U_2,U_1\right]
\equiv \left< U_2\left| e^{-\tau H}
P\right| U_1\right>
\label{keru}
\eeq
which has the invariance property 
\begin{equation}
K[\tau ;U_2,U_1]=K[\tau;uU_2u^{\dagger},vU_1v^{\dagger}]~~,
\end{equation}
As is shown in the subsection \ref{careqpcm}, this form of the heat kernel (\ref{keru})
has a natural expansion in terms of group characters,
\begin{equation}
K[\tau;U_2,U_1]~=~\sum_R~\chi_R(U_1) ~\exp\left(-e^2\tau
L C_2(R) \right)~\chi^*_R(U_2)
\label{hkchar}
\end{equation}
which, with the expression (\ref{unitary}) and (\ref{laplacian}) can
be seen to satisfy the heat equation
\begin{equation}
\left(\frac{\partial}{\partial\tau}+e^2L\Delta(G)\right) K[\tau;U_2,U_1]=0
\label{hkkk}
\end{equation}
where the gauge group Laplacian $\Delta(G)$ operates on $U_2$.  Also,
the boundary condition,
\begin{equation}
K[0,U_2,U_1]~=~\int [dg(x)]\prod_x\delta
\left(A_2(x)-A_1^g(x)\right)~=~({\rm const.})\cdot \delta_{\rm cl}( U_1,U_2)
\label{bccc}
\end{equation}
The delta function on the right-hand-side of (\ref{bccc}) is the
conjugation invariant delta function which equates the eigenvalues of
the two unitary matrices.  It can be defined by group integration,
\begin{equation}
\delta_{\rm cl}(g_1,g_2)=\int [dV]\delta( Vg_1V^{\dagger}g_2,{\cal I})
=\sum_R\chi_R(g_1)\chi_R^*(g_2)
\end{equation}

Furthermore, 
the heat kernel for 2-dimensional Yang-Mills theory on a cylinder of
length $\tau$ and base circle $L$
can be written in terms of the partition function of a gauged principal 
chiral model with open boundaries. This, as we showed in Section 2, is equivalent 
to 2D QCD if we restrict our attention to Wilson loops. The path integral 
representation of the heat kernel for that theory should have a standard form using 
action (\ref{mma}) and integration variables $A$ and $U$. 
Some care must be taken to ensure that the
heat kernel is a correct function of the eigenvalues of the Wilson
loop operators. In particular the sewing property, 
\begin{equation}
Z[\tau;U_2,U_1]=\int dU(u) Z[u;U_2,U(u)]
Z[\tau-u;U(u),U_1]\ ,
\label{sewing}
\end{equation}
must be satisfied.

 The conditions that fix the
loops at the two boundaries of the cylinder (say 
$U(0)$ and $U(\tau)$) to be $U_1$ and, up to an element of $Z_N$, $U_2$,
can be imposed introducing
the corresponding delta-functions:
\begin{eqnarray}
&&Z\left[\tau;U_1,U_2\right]=\int
\prod_{t\in [0,\tau]}^{}[dA(t)][dU(t)] e^{-\frac{1}{2e^2L}\int_0^\tau
dt{\rm 
Tr}\left| \dot U -i\left[A,U\right]\right|^2}\cr
&&\delta_{\rm cl}\left(U(0),U_1\right) \delta_{\rm
cl}\left(U(\tau),U_2\right) 
\psi(U(0))\psi(U(\tau))\ \ .
\label{kernelp}
\end{eqnarray}
The factors $\psi(U(0))$ and $\psi(U(\tau))$ are 
boundary wave functions that have to be 
introduced in order to guarantee that the sewing prescription (\ref{sewing}) 
for the kernel is satisfied.
By means of eq.(\ref{kernelp}), the left hand side of eq.(\ref{sewing})
reads
\begin{eqnarray}
& &\int dU(u)
\prod_{t\in [0,u]}^{}[dA(t)][dU(t)] \prod_{t\in
[u,\tau]}[dA'(t)][dU'(t)] 
e^{-\frac{1}{2e^2L}\int_0^u dt{\rm 
Tr}\left| \dot U
-i\left[A,U\right]\right|^2}\cr
&&e^{-\frac{1}{2e^2L}\int_u^\tau dt{\rm 
Tr}\left| \dot U' -i\left[A',U'\right]\right|^2}
\delta_{\rm cl}\left(U(0),U_1\right) 
\delta_{\rm cl}\left(U(u),
U(u)\right)\delta_{\rm cl}\left(U'(u),U(u)\right)\cr
&&\delta_{\rm cl}\left(U'(\tau),U_2\right)
\psi(U(0))\psi(U(u))
\psi(U'(u))\psi(U'(\tau))\ \ .
\end{eqnarray}
We will show in the next section that  the integration over one of the $A$
variables at the point $u$, produces
the squared inverse of the Vandermonde determinant for a unitary
matrix, $J^{-2}(U(u))$ 
\begin{equation}
J(U)=\prod_{\alpha<\beta}2\sin\frac{1}{2}\left(\phi^\alpha-
\phi^\beta\right)
\end{equation}
which only depends on the eigenvalues of $U$,
$\exp\left(i\phi^\alpha\right)$, $\alpha=1,\ldots,N$.
Consequently, Eq.(\ref{sewing}) will be satisfied if we
choose $\psi(U)=J(U)$, so that
\begin{eqnarray}
K[\tau; U_2,U_1]=\int [dA(t)][dU(t)]&~\exp\left(
-\frac{1}{2e^2L}\int_0^\tau
dt~{\rm Tr}\left| \dot U(t)-i\left[A(t),U(t)\right]\right|^2\right)\cdot
\nonumber\\
\cdot
&\delta_{\rm cl}(U(0),U_1)\delta_{\rm cl}(U(\tau),U_2) J(U(0))J(U(\tau))\ .
\label{hk4}
\end{eqnarray}

The formulae (\ref{hk3}) and (\ref{hk4}) are two alternative
expressions for the heat kernel, one in terms of matrix quantum
mechanics with a background gauge field and the other in terms of
gauge invariant matrix mechanics with the external matrices appearing
as boundary conditions.

The propagator in a $\theta$-sector can be obtained by projecting
either the initial or final state of the heat kernel onto
$\theta$-states.  This is done by summing over all transformations of
$U_1$ or $U_2$ by elements of the center of $G$, $C$. and, in that
sum, weighting each term by the phases ${\cal Z}(z,\theta)$
corresponding to the $\theta$-representation of $C$,
\begin{equation}
K[\tau,\theta;U_2,U_1]~=~\sum_{z\in C}{\cal Z}(z,\theta)K[\tau;zU_2,U_1]
\end{equation}
For the three presentations of the heat kernel given above, this has the
effect 
\begin{eqnarray}
&& K[\tau,\theta;U_2,U_1]=
\sum_R~\delta\left(R,R_\theta\right)\chi_R^*(U_2)\chi_R(U_1)
\cdot
\exp\left(-e^2L\tau C_2(R)\right)
\label{hk51}\\
&&=\sum_z{\cal Z}(z,\theta)
\int [dg_z(x)]~\exp\left(-\frac{1}{2e^2\tau}
\int_0^L dx~ {\rm Tr}\left|\nabla g_z(x)+ 
ig_z(x)A_1(x)-iA_2(x)g_z(x)\right|^2\right)
\nonumber\\\label{hk52}\\
&&=\sum_z{\cal Z}(z,\theta)
\int [dA(t)][dU(t)]~\exp\left(-\frac{1}{2e^2L}\int_0^\tau
dt~{\rm Tr}\left| \dot U(t)-i\left[A(t),U(t)\right]\right|^2\right)
\cdot
\nonumber\\
&&~~~~~~\cdot\,
\delta_{\rm cl}(U(0),U_1)\delta_{\rm cl}(U(\tau),zU_2) J(U(0))J(U(\tau))
\label{hk53}
\end{eqnarray}
Here, $R_\theta$ are those irreducible representations of $G$ which
have the property $$
\chi_{R_\theta}(zU)={\cal Z}(z,\theta)\chi_{R_\theta}(U)\ .
\label{rtheta}
$$

In the next section, by means of a diagonalization technique,
we shall show that the functional integral 
in (\ref{hk53}) can be actually performed so that the heat kernel can 
be expressed directly in terms of the eigenvalues of $U_1$ and $U_2$. 

\subsection{Calculation of the heat kernel by diagonalization technique} 
\label{kernels}

In Ref.\cite{dadda} two dimensional Yang-Mills theories were
written in terms of a Kazakov-Migdal model and the heat kernel on the
cylinder was computed using a diagonalization procedure. We shall
follow here an analogous procedure for the calculation of the
partition function (\ref{hk53}) of the gauged principal chiral model.
In fact, the integral in (\ref{hk53}) can be done by using gauge
symmetry to diagonalize at each point the unitary matrix
$U(t)=V(t)U^D(t)V^{-1}(t)$ where $U^D(t)=\hbox{\rm diag}(
e^{i\varphi^1(t)}, e^{i\varphi^2(t)},\dots,e^{i\varphi^N(t)})$.  The
$\varphi$ variables are angles since $\varphi\in[0,2\pi]$. The measure
in the integral has the form
\begin{equation}
[dU(t)]=\prod_{t\in[0,\tau]}\prod_\alpha
d\varphi^\alpha(t) J^2(\varphi(t))
\delta\left(\sum_\alpha\varphi^\alpha(t)\right)
[dV(t)]\ \ ,
\end{equation}
where $J(\varphi(t))$ is the Vandermonde determinant for a unitary
matrix,
and the action is 
\begin{equation}
S=\frac{1}{2e^2 L}\int_0^\tau dt
\left(\sum_{\alpha=1}^N \dot
\varphi^\alpha\dot\varphi^\alpha+\sum_{\alpha,\beta=1}^N\vert
A_{\alpha
\beta}\vert^2\vert
e^{i\varphi^\alpha}-e^{i\varphi^\beta}\vert^2\right)\ \ .
\end{equation}  
The integral over $A_{\alpha\beta}$, where $\alpha\neq\beta$, cancels
the Vandermonde determinant in the integration measure.  The
integration over the diagonal components of $A$ yields an infinite
factor which compensates the infinite normalization of the plane-wave
states which were used in (\ref{keru}).  The kernel (\ref{hk53})
becomes an integral over the eigenvalue variables. In the
$\theta$-sector using for $\sum_z{\cal Z}(z,\theta)$ the
representation $\sum_n\exp(-i n\theta_k)$ with $\theta_k=2\pi k/N$, we
get
\begin{eqnarray}
&&Z=\sum_{n=0}^{N-1} e^{-i\theta_k n}\int
\prod_{t\in[0,\tau]}\prod_{\alpha=1}^{N} d\varphi^\alpha(t)
\delta\left(\sum_{\alpha=1}^N\varphi^\alpha(t)\right) 
e^{-\frac{1}{2e^2 L}\int_0^\tau dt \sum_{\alpha=1}^{N}\dot
\varphi^\alpha\dot\phi^\alpha}J(\varphi(0))
J(\varphi(\tau))\cr &&\int 
dV(0)\delta\left(V(0)e^{i\varphi(0)}V^{\dag}(0)e^{-i\lambda_1},I\right)
\int 
dV(\tau)\delta\left(V(\tau)e^{i\varphi(\tau)}V^{\dag}(\tau)
e^{-i\lambda_2+2\pi in/N},I\right)\cr
&&
\label{zsun}
\end{eqnarray}
where $e^{i\lambda_{1,2}}$ denote the diagonal forms of $U_{1,2}$ and the
integration
measure integrates $\varphi(t)$, at each point t, over the range
$[0,2\pi]$.  The integral is
invariant under the field translation symmetry
\begin{equation}
\varphi^\alpha(t)
\rightarrow \varphi^\alpha(t)+2\pi n^\alpha
\label{symm}
\end{equation}
where $n^\alpha$ is an integer.\footnote{Actually in this case 
$n_\alpha$ can be any constant, but when computing loop correlators 
only the symmetry (\ref{symm}) with integer $n_\alpha$
will survive. Note that this
is not a symmetry of
the action or the integration measure in the path integral
(\ref{hk53}) separately but only appears after 
diagonalization.}
This symmetry can be used to extend the limits on the
integration over $\varphi(t)$ at each $t$ to infinity. 
 
The integrals on $\varphi^\alpha(t)$ with $t\in(0,\tau)$ can be performed 
by a $\zeta$-function regularization of the divergences. First one can
integrate on $\varphi^N(t)$ taking advantage of the
$\delta(\sum_\alpha\varphi^\alpha(t))$. Changing the functional
integration variables to the real and periodic $\tilde\varphi^\alpha(t)$,  
$\tilde\varphi^\alpha(\tau)=\tilde\varphi^\alpha(0)$, by
means of
\begin{equation}
\varphi^\alpha(t)=\left(\varphi^\alpha(\tau)-
\varphi^\alpha(0)\right)\frac{t}{\tau}
+\tilde\varphi^\alpha(t)\ \ ,
\label{boun}
\end{equation}
the integrals on the open interval $t\in(0,\tau)$, give
\begin{equation}
\exp\left\{-\frac{1}{2e^2 L
\tau}\left[\sum_{\alpha,\beta=1}^{N-1}(\varphi^\alpha(\tau)-
\varphi^\alpha(0))\Omega_{\alpha\beta}
(\varphi^\beta(\tau)-
\varphi^\beta(0))\right]\right\}\cdot Z_a \ \ ,
\label{open}
\end{equation}
where
\begin{equation}
Z_a=\frac{1}{\rm VOL ~G}\int
\prod_{t\in (0,\tau)}\prod_{\alpha=1}^{N-1} d\tilde\varphi^\alpha(t)
\exp\left(-\frac{1}{2e^2 L\tau}\int_0^\tau dt \sum_{\alpha\beta=1}^{N-1}\dot
{\tilde\varphi}^\alpha \Omega_{\alpha\beta}\dot{\tilde\varphi}^\beta\right)
\end{equation}
and $\Omega$ is the $(N-1)\times(N-1)$ matrix
\begin{equation}
\Omega=\pmatrix{2&1&\ldots&1\cr
1&2&\ldots&1\cr
\vdots&\vdots&\ddots&\vdots\cr
1&1&\ldots&2\cr}\ \ .
\label{omega}
\end{equation}
Expanding $\tilde\varphi^\alpha(t)$ in modes
\begin{equation}
\varphi^\alpha(t)={\frac{1}{\sqrt{\tau}}}\sum_{k=-\infty}^\infty
a^\alpha_k e^{i2\pi k t/\tau} \,
\end{equation}
the functional measure is defined as $\prod_td\varphi(t)\equiv
\prod_k da_k$.  The integration over the zero modes produce an
infinite, irrelevant, temperature independent factor proportional to
the volume of the gauge group.  The functional integral in $Z_a$ is
proportional to the determinant of the Laplacian,
\begin{equation}
Z_a=\prod_{k\neq 0}\left(\det \frac{e^2 L \tau}{2\pi}\Omega
k^2\right)^{-\frac{1}{2}}
=\det\left( \frac{e^2 L\tau}{2\pi}\Omega\right)^{-\zeta(0)}e^{(N-1)\zeta'(0)}=
\sqrt{N}\left(\frac{1}{e^2L \tau}\right)^{\frac{N-1}{2}}\ \ .
\label{za}
\end{equation}
where we have used zeta-function regularization with
\begin{equation}
\zeta(s)=\sum_{k=1}^{\infty}\frac{1}{k^{2s}}
\end{equation}
and $\zeta(0)=-1/2$ and $\zeta'(0)=-(\log 2\pi)/2$.

We are now left with the integration on the $\varphi^\alpha(t)$,
$\alpha=1,\dots,N$ at the boundaries $t=0,\tau$. This can be easily done
by rewriting the conjugate invariant delta functions in (\ref{zsun}) 
according to \cite{dadda,kmsw}
\begin{equation}
\int dV\delta\left(V e^{i\varphi} V^{\dag} 
e^{-i\lambda}\right)=\sum_P
\sum_{\{n_\alpha\}=-\infty}^{+\infty}\frac{(-1)^{P+(N-1)
\sum_{\alpha=1}^N n_\alpha}}{J(\varphi)
J(\lambda)}\prod_{\alpha=1}^N\delta(\varphi_\alpha-\lambda_{P(\alpha)}
+2\pi n_\alpha)\ \ ,
\label{delta}
\end{equation}
where $P$ denotes a permutation of the indices and $(-1)^P$ its parity.
By substituting Eqs.(\ref{open}), (\ref{za}), (\ref{delta}) into
Eq.(\ref{zsun}) and integrating in $\varphi^\alpha(0),\
\varphi^\alpha(\tau)$ we finally get
\begin{eqnarray}
&&K[\tau,\theta_k;U_2,U_1]= N!\sqrt{N}
\left(\frac{1}{e^2 L\tau}\right)^{\frac{N-1}{2}}\sum_{n=0}^{N-1}\sum_P
\frac{(-1)^P e^{-2\pi i n[k-N(N-1)/2]/N}}{J(\lambda_1)
J(\lambda_2)}\cr
&&\sum_{\{n_\alpha\}=-\infty}^{+\infty}
\delta\left(\sum_{\alpha=1}^N n_\alpha+n\right)
\exp\left[-\frac{1}{2e^2
L\tau}\sum_{\alpha=1}^N\left(\lambda_1^\alpha-\lambda_2^{P(\alpha)}+2\pi
n_\alpha+\frac{2\pi n}{N}\right)^2\right]\ \ .
\label{kercyl}
\end{eqnarray}
The delta-function imposing the constraint $\sum_\alpha
n_\alpha +n=0$ arises from the $\delta(\sum_\alpha \varphi_\alpha)$ by
taking into
account that the $U_2$ eigenvalues are twisted by an element of the center and 
that, being the gauge group $SU(N)$, we can choose 
$\sum_\alpha\lambda_\alpha=0$ for the loops at the boundaries.
We shall show in the nex subsection that Eq.(\ref{kercyl}) is equivalent to the 
character representation (\ref{hk51}).

\subsection{From the functional integral to the character expansion}
\label{careqpcm}

Following Ref.\cite{dadda}, it is possible to show that,
by means of a Poisson resummation, Eq.(\ref{kercyl}) can be rewritten
in terms of a character expansion. It is convenient first to rewrite the 
delta function on the integers $n_\alpha$ as an integral
\begin{equation}
\delta\left(\sum_{\alpha=1}^{N}+n\right)=\int_0^{2\pi}\frac{d\vartheta}{2\pi}
\exp\left(i\vartheta\sum_{\alpha=1}^N n_\alpha+n\right)\ .
\label{deltain}
\end{equation}
One can then complete the square in $n_\alpha$, so as to rewrite
Eq.(\ref{kercyl}), according to
\begin{eqnarray}
&&K[\tau,\theta_k;U_1,U_2]= N!
\left(\frac{1}{e^2 L\tau}\right)^{\frac{N-1}{2}}\sum_{n=0}^{N-1}\sum_P
\frac{(-1)^P e^{-2\pi i n[k-N(N-1)/2]/N}}{J(\lambda_1)
J(\lambda_2)}\cr
&&\sum_{\{n_\alpha\}=-\infty}^{+\infty}
\int_0^{2\pi}\frac{d\vartheta}{2\pi}
\exp\left[-\frac{2\pi^2}{e^2
L\tau}\sum_{\alpha=1}^N\left(\frac{\lambda_1^\alpha-\lambda_2^{P(\alpha)}}{2\pi}+
n_\alpha+\frac{n}{N}-\frac{i\vartheta L\tau e^2}{4\pi^2 
}\right)^2-\frac{N\vartheta^2 e^2 L\tau}{8\pi^2}\right]\ \ ,
\cr &&
\label{prepois}
\end{eqnarray}
where we used that $\sum_{\alpha=1}^N\lambda^i_\alpha=0$, $i=1,2$.
Applying
the generalized Poisson resummation formula
\begin{equation}
\sum_{\vec m=-\infty}^{+\infty}\exp\left(-\pi g_{ij}m^i m^j -2\pi i
m^ia_i\right)=\frac{1}{\sqrt{\det g_{ij}}}
\sum_{\vec n=-\infty}^{+\infty}\exp\left[-\pi
g^{ij}(n_i-a_i)(n_j-a_j)\right]\ . \label{pois}
\end{equation}
where $g^{ij}$ is $g_{ij}$ inverse, and then completing  the square in
$\vartheta$, Eq.(\ref{prepois})
becomes
\begin{eqnarray}
&&K[\tau,\theta_k;U_1,U_2]=
\left(e^2 L\tau N\right)^{1/2}\sum_{\{m_\alpha\}}
\sum_{n=0}^{N-1}\sum_{P,P'}
\frac{(-1)^{(P+P')} e^{-2\pi i n[k-N(N-1)/2]/N}}{J(\lambda_1)
J(\lambda_2)}\cr
&&\exp\left[i\sum_{\alpha=1}^N m_\alpha\left(\lambda_1^{P(\alpha)}
-\lambda_2^{P'(\alpha)}+2\pi i n\frac{1}{N}\right)\right]
\exp\left\{-\frac{e^2 L\tau}{2}\left[\sum_{\alpha=1}^N
m_\alpha^2-\frac{1}{N}\left(\sum_{\alpha=1}^N
m_\alpha\right)^2\right]\right\}\cr
&&\left(\frac{1}{2\pi}\right)^{(1+N/2)}
\int_0^{2\pi} d\vartheta\exp\left[-\frac{Ne^2L\tau}{8\pi^2}
\left(\vartheta-\frac{2\pi}{N}\sum_{\alpha=1}^Nm_\alpha\right)^2\right]\
\ ,
\label{malpha}
\end{eqnarray}
where we introduced a redundant sum over permutation canceling the $N!$
factor.
Note that Eq.(\ref{malpha}) is symmetric with respect to any permutation
of the $m_\alpha$ and vanishes when any two of $m_\alpha$, $m_\beta$ are
equal. Therefore one can introduce an ordering between the integers
$m_1>m_2>\dots>m_N$ multiplying by a factor $N!$. One can further notice
that all the terms, except the last integral are invariant if all the
$m_\alpha$ are shifted by the same integer. One can then define the
integers $r_\alpha=m_\alpha-m_N -N$, $\alpha=1,\dots,N$ and use the sum
over $m_N$ to complete the Gaussian integral in $\vartheta$ by shifting
$\vartheta\to\vartheta-2\pi m_N - 2\pi N$. These
manipulations lead to
\begin{eqnarray}
&&K[\tau,\theta_k;U_1,U_2]=
N!\sum_{n=0}^{N-1}
\sum_{r_1>r_2\dots>r_N=-N}
\exp\left\{-\frac{2\pi i n}{N}\left[k-\frac{N(N-1)}{2}-\sum_{\alpha=1}^N
r_\alpha\right]\right\}\cr
&&\left(\frac{1}{2\pi}\right)^{(N-1)/2} 
\frac{\det e^{i r_\alpha\lambda^1_\beta}}{J(\lambda_1)}
\frac{\det e^{-i r_\alpha\lambda^2_\beta}}{J(\lambda_2)}
\exp\left\{-\frac{e^2 L\tau }{2}\left[\sum_{\alpha=1}^N
r_\alpha^2-\frac{1}{N}\left(\sum_{\alpha=1}^N
r_\alpha\right)^2\right]\right\}
\label{almost}
\end{eqnarray}
The ingredients of Eq.(\ref{almost}) are now easy to recognize. The last
exponent is, up to a constant, the $SU(N)$ quadratic Casimir in the
representation $R$ whose
Young table has $r_\alpha+\alpha$ boxes in the $\alpha^{\rm th}$ row.
\begin{equation}
\sum_{\alpha=1}^N
r_\alpha^2-\frac{1}{N}\left(\sum_{\alpha=1}^N
r_\alpha\right)^2=2 C_2(R)+\frac{N(N^2-1)}{12}\ .
\label{casi}
\end{equation}
The second term in the r.h.s. of Eq.(\ref{casi}) is the trace of the
inverse Cartan matrix and is related to
the scalar curvature of the group manifold. It provides the energy of the
lowest representation.
The first exponential in Eq.(\ref{almost}) projects the sum to those
representations $R_k$ whose
Young tableaux have rows with a number of boxes equal to the theta vacuum
$k$, ${\rm mod}[N]$. Moreover the determinants are related to the
character of the matrix
$U_{1,2}$ in the representation $R$.
\begin{equation}
\chi_R(U)=\frac{\det\left(e^{i
r_\alpha\lambda_\beta}\right)}{J(\lambda)}  \ .
\end{equation}
Up to an irrelevant, temperature independent constant, one can then write
the final equation as
\begin{equation}
K[\tau,\theta_k;U_1,U_2]=\sum_{R_k} \chi_{R_k}(U_1) \chi^*_{R_k}(U_2)
\exp\left\{-\frac{e^2
L\tau}{2}\left[2 C_2(R_k)+\frac{N(N^2-1)}{12}\right]\right\}\ \ .
\end{equation}

\subsection{2D Yang-Mills theory on the sphere as matrix quantum mechanics}

Once one has the heat kernel for 2D Yang-Mills it is an easy matter to
construct the partition function for the theory on a sphere or a
torus.  If one imagines shrinking the ends of the cylinder to points
then the resulting topology is that of a sphere.  In terms of the the
discussion of the previous section this corresponds to setting
$U_1=U_2=1$.  Consequently, from (\ref{hkchar}) the partition function
on the sphere is
\begin{equation}
Z[S^2;\tau,\theta]=K[\tau;\theta,1,1]=\sum_R 
\delta\left(R,R_\theta\right)(\rm{dim}(R))^2 \exp\left(-e^2\tau L
C_2(R) \right)
\end{equation}
where we have used the fact that $\chi_R(1)= \rm{dim}(R)$ and $R_\theta$ is 
defined in (\ref{rtheta}).

More interestingly, if we carry out the equivalent procedure with the
path integral principal chiral model of the heat kernel (\ref{hk3})
and set $A_1(x) = A_2(x) =0$ we find the following form
\beq
Z[S^2;\tau,\theta]=\sum_z{\cal Z}(z,\theta)\int 
[dg_z(x)]~\exp\left( -\frac{1}{2e^2\tau}
\int_0^L dx {\rm Tr}\left|\nabla g_z(x) \right|^2\right)
\eeq
where $g(L)=zg(0)$.

Consequently we can interpret the partition function for the sphere as
unitary matrix quantum mechanics.  

\subsection{2D Yang-Mills theory on the torus as gauge invariant 
matrix quantum mechanics}

Similar to the case of the sphere it is easy to imagine manipulating
a cylinder to form a torus - identifying the ends and sewing them
together - which produces the trace of the heat kernel.  This process
is easy to carry out in the character representation by identifying
$U_1=U_2$ and integrating. Hence we have the partition function on a
torus of area $\tau L$ 
\beq
Z[T^2;\tau,\theta]=\int[dU]~K[\tau,\theta;U,U]
=~\sum_R\delta\left( R,R_{\theta}
\right)\exp\left( -e^2\tau L C_2(R)\right) \eeq
 
Explicitly, for the special case of SU(N), where representations are
denoted by the usual Young Tableau row variables $l_1\ge l_2 \ge
\cdots \ge l_N \ge 0$ with $l= \sum l_j$ we can give an explicit form
for the partition function on the torus.  When $\theta=2\pi k/N$, we
consider only those representations where $\sum_i l_i= k~({\rm
  mod}N)$.  Then 
\beq Z[\tau,\theta_k=2\pi k/N]~=~\sum_{\{l_j \} }
\delta_N(\sum_i l_i,k) \exp\left[ -e^2\tau L \sum_{j=1}^N l_j ( l_j +
  N + 1 - 2 j -l/N) \right] \eeq 

Equivalently, in the following section, we shall show 
how this result can be obtained by taking 
the trace of the heat kernel represented as a functional
integral for the matrix model (\ref{kercyl}).

\subsection{Partition function on the torus in the functional integral approach}

From the kernel on the cylinder obtained in the section \ref{kernels},
Eq.(\ref{kercyl}), the partition function on the
torus can be readily obtained by sewing together
the two ends of the cylinder. Namely one takes the trace of the kernel on
the cylinder with the appropriate measure
$\prod_\alpha
d\lambda_\alpha\delta(\sum_\alpha\lambda_\alpha)J^2(\lambda)$
\begin{eqnarray}
&&Z[\tau,\theta_k]=\int dU K[\tau,\theta_k;U,U]
=N!\sqrt{N}\left(\frac{1}{e^2
L\tau}\right)^{\frac{N-1}{2}}\sum_{n=0}^{N-1}\sum_P
(-1)^P \cr
&&\exp\left\{-2\pi i n\frac{1}{N}[k-\frac{N(N-1)}{2}]\right\}
\sum_{\{n_\alpha\}=-\infty}^{+\infty}
\delta\left(\sum_{\alpha=1}^N n_\alpha+n\right)
\int_0^{2\pi}\prod^N_{\alpha=1}
d\lambda_\alpha\delta\left(\sum_{\alpha=1}^N
\lambda_\alpha\right)\cr
&&\exp\left[-\frac{1}{2e^2
L\tau}\sum_{\alpha=1}^N\left(\lambda^1_\alpha-\lambda^2_{P(\alpha)}+2\pi
n_\alpha+\frac{2\pi n}{N}\right)^2\right]\ \ .
\cr&&
\label{ztor}
\end{eqnarray}
To calculate the integral in (\ref{ztor}) we can proceed as in the
previous section, first writing the $\delta(\sum_\alpha n_\alpha+n)$ as an
integral, completing the square in $n_\alpha$ and then Poisson resumming.
One gets
\begin{eqnarray}
&&Z[\tau,\theta_k]=N!\left(\frac{1}{2\pi}\right)^{1+\frac{N}{2}}
\left(Ne^2 L\tau\right)^{\frac{1}{2}}\sum_{n=0}^{N-1}\sum_P
(-1)^P\sum_{\{n_\alpha\}=-\infty}^{+\infty}
e^{-2\pi i
n\frac{1}{N}\left[k-\frac{N(N-1)}{2}-\sum_\alpha
n_\alpha\right]}\cr
&&\int_0^{2\pi}\prod^N_{\alpha=1}
d\lambda_\alpha\delta\left(\sum_{\alpha=1}^N \lambda_\alpha\right)
\exp\left\{-\frac{e^2 L\tau}{2}
\left[\sum_{\alpha=1}^N n_\alpha^2-\frac{1}{N}\left(\sum^N_{\alpha=1}
n_\alpha\right)^2\right]+i\sum_{\alpha=1}^N
(n_\alpha-n_{P(\alpha)})\lambda_\alpha\right\}\cr
&&\int_0^{2\pi} d\vartheta \exp\left[-\frac{Ne^2 L\tau}{8\pi^2}
\left(\vartheta-\frac{2\pi}{N}\sum_{\alpha=1}^N
n_\alpha\right)^2\right] \ \ .
\label{ztoro}
\end{eqnarray}
We can now shift the integer $n_\alpha\to n_\alpha -n_N-N=r_\alpha$
$\alpha=1,\dots,N$ and use the sum over $n_N$ to complete the
Gaussian integral in $\vartheta$.
The $\delta(\sum_\alpha\lambda_\alpha)$ can be eliminated integrating in 
$\lambda_N$. The integrals in $\lambda_\alpha$, $\alpha=1,\dots,N-1$ give
just a product of Kronecker deltas imposing the $N-1$ conditions
\begin{equation}
r_\alpha-r_N -r_{P(\alpha)} + r_{P(N)}=0\
,~~~~~~~~~~~~~~\alpha=1,\dots,N-1\ ,
\label{krondel}
\end{equation}
these can actually be extended to $\alpha=N$, because the $N^{\rm
th}$ is trivially satisfyed. The sum of the conditions (\ref{krondel}),
for $\alpha=1,\dots,N$,
gives $r_N-r_{P(N)}=0$, so that the product of Kronecker give rise to the
determinant factor
\begin{equation}
\sum_P (-1)^P
\prod^N_{\alpha=1}\delta_{r_{P(\alpha)},r_\alpha}=\det_{\alpha\beta}
\delta_{r_\alpha r_\beta}\ \ . 
\end{equation} 
The $SU(N)$ partition function on the torus then reads
\begin{eqnarray}
&&Z[\tau,\theta_k]=N!(2\pi)^{(N-1)/2}
\sum_{n=0}^{N-1}\sum_{\{r_1,\dots,r_{N-1}\}=-\infty}^{+\infty}
\exp\left\{-2\pi i
n\frac{1}{N}\left[k-\frac{N(N-1)}{2}-\sum_{\alpha=1}^N
r_\alpha\right]\right\}\cr
&&\det_{\alpha\beta}\left[\delta_{r_\alpha,r_{\beta}}\right]
\exp\left\{-\frac{e^2 L\tau}{2}
\left[\sum_{\alpha=1}^N r_\alpha^2-\frac{1}{N}\left(\sum^N_{\alpha=1}
r_\alpha\right)^2\right]\right\}\ \ .
\label{ztorus}
\end{eqnarray}
The determinant of Kronecker deltas forbids any two of the integers
$r_\alpha$ $\alpha=1,\dots,N$ from taking the same value and, the
expression (\ref{ztorus}) being
symmetric under the permutation of any $r_\alpha$, the integers can be
ordered according to $r_1>\dots>r_N=-N$ provided we multiply by a factor
$N!$. 

The result (\ref{ztorus}) gives the correct spectrum for 2-dimensional
$SU(N)$ Yang-Mills theories on the torus which, without the theta
angle, has already been provided in many papers \cite{rus,wit,blau};
it does not agree with the one found in ref.\cite{wipf}.
$Z_{SU(N)}$ is a sum over the representation of the exponential of the
second Casimir, sum which is actually restricted only to those
representations whose Young tableaux have rows with $k$ ${\rm mod}[N]$
boxes, where $k$ is the discrete theta angle of 2-dimensional
Yang-Mills theories.  To compute (\ref{ztorus}) we followed the path
integral calculation in \cite{dadda} generalizing it to the case of
non trivial theta states.   

\section{Loop Correlators on the Cylinder}
\setcounter{equation}{0}

In this section we shall  compute, on the cylinder, the correlator of
two Wilson loops in the fundamental representation. One loop is situated  at the point $u$ with 
$l$ windings and and the other, at the point $v$ ,with $m$.
\begin{equation}
P_{l,m}(\theta_k, U_1,U_2;u,v)=\frac{1}{N^2 Z_k}\left<U_1\left|{\rm
Tr}\{U^l(u)\} 
{\rm Tr}\{U^m(v)\}P_k\right|U_2\right>\ \ ,
\label{corr}
\end{equation}
where $Z_k$ coincides with the kernel on the cylinder in the $k^{\rm th}$ 
$\theta$-sector ($e.g.$ Eq.(\ref{kercyl})).
This will allow us to obtain also the correlator on the torus just by
sewing together the two ends of the cylinder, as we did for the partition
function in the previous section.
Using the prescriptions introduced for the partition function on the
cylinder, the path integral representation 
of the correlator (\ref{corr}) reads
\begin{eqnarray}
&&P_{l,m}(\theta_k,U_1,U_2;u,v)=\frac{1}{N^2 Z_k}
\sum_{n} e^{-in\theta_k}\int
\prod_{t\in [0,\tau]}^{}[dA(t)][dU(t)] e^{-\frac{1}{2e^2 L}\int_0^\tau
dt{\rm 
Tr}\left| \dot U -i\left[A,U\right]\right|^2}\cr
&&{\rm Tr}\{U^l(u)\}{\rm Tr}\{U^m(v)\}\delta_{\rm
cl}\left(U(0),U_1\right) \delta_{\rm
cl}\left(U(\tau),e^{2\pi i n/N}U_2\right) 
\psi(U(0))\psi(U(\tau))\ \ .
\label{corp}
\end{eqnarray}
Taking, for the time being, $v>u$, in order to compute (\ref{corp}) it is
most convenient to perform first the integration
in the open intervals 
$t\in (0,u)$, $(u,v)$ and $(v,\tau)$. 
This corresponds to the calculation  of the partition function on
the three cylinders with boundaries $(U_1,U(u))$, $(U(u),U(v))$ and
$(U(v),U_2)$.  After the diagonalization and the integration on the
gauge potential in the whole interval $t\in [0,\tau]$, we can perform the 
path integral in
the open intervals $t\in (0,u)$, $(u,v)$ and $(v,\tau)$ 
as we did in  the previous section
in the open interval $t\in(0,\tau)$.
The results will then be given by Eqs.(\ref{open}) and (\ref{za})
with the appropriate changes for the different lengths of the three
cylinders. Then, using Eq.(\ref{delta}) for the delta-functions in
$(\ref{corp})$ one can integrate on the boundary points $t=0,\
t=\tau$. The result reads
\begin{eqnarray}
P_{l,m}&& \!\!\!\!\!\!\!\!\!\!(\theta_k,\lambda_1,\lambda_2;u,v) =  \label{prepoisco}
\\ &&
\frac{1}{\sqrt{N} Z_k}
\left(\frac{1}{(e^2L)^3
u(v-u)(\tau-v)}\right)^{\frac{N-1}{2}}\sum_{n=0}^{N-1}\sum_{P,P'}
\frac{(-1)^{P+P'} e^{-2\pi i n[k-N(N-1)/2]/N}}{J(\lambda_1)
J(\lambda_2)}
\cr
\nonumber &&
\sum_{\{n_\alpha,l_\alpha\}=-\infty}^{+\infty}
\delta\left(\sum_{\alpha=1}^N n_\alpha\right)
\delta\left(\sum_{\alpha=1}^N l_\alpha+n\right)
\int\prod_{\alpha=1}^N
d\varphi^\alpha(u)d\varphi^\alpha(v)\sum_{\beta,\gamma=1}^N
e^{il\varphi^\beta(u)+im\varphi^\gamma(v)} 
\cr &&
\delta\left(\sum_{\alpha=1}^N\varphi^\alpha(u)\right) 
\delta\left(\sum_{\alpha=1}^N\varphi^\alpha(v)\right)
\exp\left[-\frac{1}{2e^2 L
u}\sum_{\alpha=1}^N\left(\lambda_1^\alpha-\varphi^{P(\alpha)}(u)+2\pi
n_\alpha\right)^2\right]
\cr &&
\exp\left[-\frac{1}{2e^2 L
(v-u)}\sum_{\alpha=1}^N\left(\varphi^\alpha(u) -\varphi^{\alpha}(v)\right)^2 \right.   \\
&& \left.  \;\;\;\;\;\;\;\;\;\;\;\;\;\;\;\;\;\;\;\;\;\;\;\;
- \frac{1}{2e^2 L
(\tau-v)}\sum_{\alpha=1}^N\left(\varphi^\alpha 
(v)-\lambda_2^{P'(\alpha)}+2\pi
l_\alpha+2\pi\frac{n}{N}\right)^2\right] \nonumber \ , 
\end{eqnarray}
where we have used the delta-functions in (\ref{prepoisco}) to eliminate
the matrix $\Omega$ appearing in (\ref{open}).
We can now proceed as before: write the delta-functions in the integers
as in Eq.(\ref{deltain}), perform a generalized Poisson resummation
in $n_\alpha$ and $l_\alpha$ to the new integers $n^{\prime}_\alpha$ and
$l^{\prime}_\alpha$, introduce the integers
$r_\alpha=n^{\prime}_\alpha-n^{\prime}_N-N$ and
$s_\alpha=l^{\prime}_\alpha-l^{\prime}_N-N$ ($r_N=s_N=-N$)
and integrate on
the variables introduced to impose the constraints on the integers (say
$\vartheta$ and $\vartheta'$) using
the sums over
$n^{\prime}_N$ and $l^{\prime}_N$ to
complete the Gaussian integrals in $\vartheta$ and $\vartheta'$.
\begin{eqnarray}
&&P_{l,m} (\theta_k,\lambda_1,\lambda_2;u,v)=  \label{postpois} \\  && \frac{1}{N^{3/2} Z_k}
\left(\frac{1}{4e^2 L \pi^2
(v-u)}\right)^{\frac{N-1}{2}}\sum_{n=0}^{N-1}
\sum_{\{r_\alpha,s_\alpha\}=-\infty}^{+\infty}
\sum_{P,P'}(-1)^{P+P'}
\frac{e^{-2\pi i n[k-N(N-1)/2-\sum_\alpha
s_\alpha]/N}}{J(\lambda_1)
J(\lambda_2)}\cr
&&
\exp\left\{-\frac{e^2 L
u}{2}\left[\sum_{\alpha=1}^N r_\alpha^2-\frac{1}{N}
\left(\sum_{\alpha=1}^N r_\alpha\right)^2\right]
-\frac{e^2 L(\tau-v)}{2}\left[\sum_{\alpha=1}^N s_\alpha^2-\frac{1}{N}
\left(\sum_{\alpha=1}^N
s_\alpha\right)^2\right]\right\} \cr&&
\exp\left[i\sum_{\alpha=0}^N\left(r_\alpha\lambda_1^{P(\alpha)}
-s_\alpha\lambda_2^{P'(\alpha)}\right)\right]
\int\prod_{\alpha=1}^N
d\varphi^\alpha(u)d\varphi^\alpha(v)\sum_{\beta,\gamma=1}^N
\delta\left(\sum_{\alpha=1}^N\varphi^\alpha(u)\right)
\delta\left(\sum_{\alpha=1}^N\varphi^\alpha(v)\right)\cr
&&\exp\left[il\varphi^\beta(u)+im\varphi^\gamma(v)-
i\sum_{\alpha=0}^N\left(r_\alpha\varphi^{\alpha}(u)
-s_\alpha\varphi^{\alpha}(v)\right)
-\frac{1}{2 e^2 L
(v-u)}\sum_{\alpha=1}^N\left(\varphi^\alpha(u)-
\varphi^{\alpha}(v)\right)^2\right]
\nonumber
\end{eqnarray}
Taking advantage of the delta-functions in Eq.(\ref{postpois}) to
integrate on $\varphi^N(u)$ and $\varphi^N(v)$ the integral 
in (\ref{postpois}) becomes
\begin{eqnarray}
&&\int\prod_{\alpha=1}^{N-1}
d\varphi^\alpha(u)d\varphi^\alpha(v)\sum_{\beta,\gamma=1}^N
\exp\left[-i\sum_{\alpha=0}^{N-1}\left(r_\alpha-r_N -
l(\delta_{\alpha\beta}-\delta_{N\beta})\right)\varphi^{\alpha}(u)
\right]\cr
&&\exp\left[i\sum_{\alpha=0}^{N-1}\left(s_\alpha-s_N -
m(\delta_{\alpha\gamma}-\delta_{N\gamma})\right)
\varphi^{\alpha}(v)\right]\cr
&&\exp\left[-\frac{1}{2e^2 L(v-u)}\sum_{\alpha\delta=1}^{N-1}
\left(\varphi^\alpha(u)-
\varphi^{\alpha}(v)\right)\Omega_{\alpha\delta}
\left(\varphi^\delta(u)-
\varphi^{\delta}(v)\right)\right]\ .
\end{eqnarray}
We can now complete the square, for example in $\varphi^\alpha(u)$,
using the inverse of the matrix $\Omega$,
\begin{equation}
\Omega^{-1}_{\alpha\beta}=\frac{1}{N}\pmatrix{N-1&-1&\ldots&-1\cr
-1&N-1&\ldots&-1\cr
\vdots&\vdots&\ddots&\vdots\cr
-1&-1&\ldots&N-1\cr} \ \ ,
\label{ginv}
\end{equation}
and then perform the Gaussian integration. 
At this point the integrals on
the $\varphi^\alpha(v)$ just give products of Kronecker deltas relating
the integers $s_\alpha$ to the $r_\alpha$ through the conditions
\begin{equation}
s_\alpha=r_\alpha-n(\delta_{\alpha\beta}-\delta_{N\beta})-
m(\delta_{\alpha\gamma}-\delta_{N\gamma})\ \ .
\end{equation}
Summing on the $s_\alpha$, we finally get, for the $SU(N)$ correlator of
two loops on a cylinder ($v>u$)
\begin{eqnarray}
&&P_{l,m}(\theta_k,\lambda_1,\lambda_2;u,v)=  \frac{1}{N^2 Z_k}
\left(\frac{1}{2\pi}\right)^{\frac{N-1}{2}}
\sum_{\{r_1,\dots,r_{N-1}\}=-\infty}^{+\infty}
\sum_{n=0}^{N-1}
e^{-2\pi in\frac{1}{N}
[k+l+m-N(N-1)/2-\sum_\alpha r_\alpha]}\cr
&&\frac{1}{J(\lambda_1)J(\lambda_2)}\sum_{\beta,\gamma=1}^N
\sum_{P,P'}(-1)^{P+P'} 
\exp\left[i\sum_{\alpha=0}^N\left(r_\alpha\lambda_1^{P(\alpha)}
-(r_\alpha-l\delta_{\alpha\beta}-m\delta_{\alpha\gamma})
\lambda_2^{P'(\alpha)}\right)\right]
\cr
&&\exp\left\{-\frac{e^2
L\tau}{2}\left[\sum_{\alpha=1}^N r_\alpha^2-\frac{1}{N}
\left(\sum_{\alpha=1}^N r_\alpha\right)^2\right]
-\frac{e^2 L(\tau-u)}{2}\left[l^2\left(\frac{N-1}{N}\right)+
\frac{2l}{N}\sum_{\alpha=1}^N
r_\alpha-2lr_\beta\right]\right\}\cr
&&
\exp\left\{-\frac{e^2 L(\tau-v)}{2}\left[m^2\left(\frac{N-1}{N}\right)+
\frac{2m}{N}\sum_{\alpha=1}^N
r_\alpha-2mr_\gamma+2lm\left(\delta_{\beta\gamma}
-\frac{1}{N}\right)\right]\right\}\ ,
\label{corcyl}
\end{eqnarray}
the case $v<u$ can be obtained by exchanging $u\leftrightarrow v$. For
$l=m=0$ we get, as we should, $1$.
Even if in (\ref{corcyl}) the sum over the integers $r_\alpha$ is
unrestricted, the presence of the determinant
$\sum_P(-1)^P\exp(i\sum_\alpha r_\alpha\lambda_1^{P(\alpha)})$ forbids any
couple of the $r_\alpha$ to take the same values. Consequently, one can
easily reconstruct in (\ref{corcyl}) the second Casimir and the sum over
representation as in Eq.(\ref{almost}). The sum will then be restricted
only to those representations whose Young tableaux have rows with a number
of boxes given by $k+l+m=0,\ {\rm mod}[N]$ where $k$ is the theta
angle and $l,\ m$ are the number of windings of the Wilson loops.

\subsection{Wilson and Polyakov loop correlators on the torus}

From Eq.(\ref{corcyl}) we can easily get the correlator of two Wilson 
(and, by duality $\tau\leftrightarrow L$, Polyakov) 
loops in the fundamental representation on the torus, just by taking the
trace of Eq.(\ref{corcyl}) according to
\begin{equation}
P_{l,m}(\theta_k;u,v)=\int \prod_{\alpha=1}^N d\lambda_\alpha
\delta\left(\sum_{\alpha=1}^N\lambda_\alpha\right)
P_{l,m}(\theta_k,\lambda,\lambda;u,v)\ .
\label{corto}
\end{equation}
Using the delta-function to eliminate the $N^{\rm th}$ component of
$\lambda$, the integrations just give a product of Kronecker deltas
imposing the conditions
\begin{equation}
r_\alpha-r_{P(\alpha)}+r_{P(N)} -r_N
-l\left(\delta_{\alpha\beta}-\delta_{N\beta}\right)
-m\left(\delta_{\alpha\gamma}-\delta_{N\gamma}\right)=0~~~~
\alpha=1,\dots,N\ ,
\label{krondelco}
\end{equation}
where the $N^{\rm th}$ equation does not arise from the integration but
can be added because is automatically satisfied.
The sum, from $1$ to $N$, of these conditions gives
\begin{equation}
r_{P(N)} -r_N -\frac{1}{N}(m+l)+l\delta_{N\beta}+m\delta_{N\gamma}=0\ ,
\end{equation}
consequently $m+l$ must be an integer multiple of $N$ or must be zero
otherwise the delta imposing the conditions (\ref{krondelco}) will set to
zero the correlator. This is nothing but the $Z_N$ invariance: all the
correlators that are not $Z_N$-invariant must vanish.
The correlator (\ref{corto}) is 
\begin{eqnarray}
&&P_{l,m}(\theta_k;u,v)=  \frac{N!}{N^2 Z_k}
(2\pi)^{\frac{N-1}{2}}
\sum_{\{r_1,\dots,r_{N-1}\}=-\infty}^{+\infty}
\sum_{n=0}^{N-1}
e^{-2\pi i n\frac{1}{N}
[l+m+k-N(N-1)/2-\sum_\alpha r_\alpha]}\cr
&&\sum_{\beta,\gamma=1}^N\sum_{P}(-1)^{P}\prod_{\alpha=1}^N
\delta_{r_{P(\alpha)},r_\alpha-l\delta_{\alpha\beta}-m\delta_{\alpha\gamma}
+(m+l)/N}\cr
&&\exp\left\{-\frac{e^2
L\tau}{2}\left[\sum_{\alpha=1}^N r_\alpha^2-\frac{1}{N}
\left(\sum_{\alpha=1}^N r_\alpha\right)^2\right]
-\frac{e^2 L(\tau-u)}{2}\left[l^2\left(\frac{N-1}{N}\right)+
\frac{2l}{N}\sum_{\alpha=1}^N
r_\alpha-2lr_\beta\right]\right\}\cr
&&
\exp\left\{-\frac{e^2 L(\tau-v)}{2}\left[m^2\left(\frac{N-1}{N}\right)+
\frac{2m}{N}\sum_{\alpha=1}^N
r_\alpha-2mr_\gamma+2lm\left(\delta_{\beta\gamma}
-\frac{1}{N}\right)\right]\right\}\ .
\label{cortoro}
\end{eqnarray}
Setting $m=0$ from Eq.(\ref{cortoro}) one gets the correlator for a
single loop. From $Z_N$ invariance, this will be different from
zero only if $l$ is an
integer multiple of $N$.
\begin{eqnarray}
&&P_{l}(\theta_k;u)=\frac{1}{N Z_k}\left<{\rm Tr}\{U^l(u)\} 
P_k\right>=\frac{N!}{N Z_k}
(2\pi)^{\frac{N-1}{2}}\cdot\cr
&&\cdot
\sum_{\{r_1,\dots,r_{N-1}\}=-\infty}^{+\infty}
\sum_{n=0}^{N-1}
e^{-2\pi i n\frac{1}{N}
[l+k-N(N-1)/2-\sum_\alpha r_\alpha]}
\sum_{\beta=1}^N\sum_{P}(-1)^{P}\prod_{\alpha=1}^N
\delta_{r_{P(\alpha)},r_\alpha-l\delta_{\alpha\beta}+l/N}\cr
&&\exp\left\{-\frac{e^2
L\tau}{2}\left[\sum_{\alpha=1}^N r_\alpha^2-\frac{1}{N}
\left(\sum_{\alpha=1}^N r_\alpha\right)^2\right]
-\frac{2 e^2 L(\tau-u)}{2}\left[l^2\left(\frac{N-1}{N}\right)+
\frac{2l}{N}\sum_{\alpha=1}^N
r_\alpha-2lr_\beta\right]\right\}\ .
\cr&&
\label{cortoro1}
\end{eqnarray}
For a loop anti-loop correlator, $m=-l$,  
Eq.(\ref{cortoro}) can be rewritten in a different form to
make clear its invariance under translations on the torus
\begin{eqnarray}
&&P_{l,-l}(\theta_k;u,v)=  \frac{(N-1)!}{Z_k}
(2\pi)^{\frac{N-1}{2}}
\sum_{\{r_1,\dots,r_{N-1}\}=-\infty}^{+\infty}
\sum_{n=0}^{N-1}
e^{-2\pi i n \frac{1}{N}
[k-N(N-1)/2-\sum_\alpha r_\alpha]}\cr
&&\left(\sum_{\beta=1}^N-\sum_{\beta,\gamma=1,\beta\ne\gamma}^N
\delta_{r_\gamma,r_\beta-l}\right)
\det_{i,j}\delta_{r_i,r_j}
\exp\left\{-\frac{e^2 L\tau}{2}\left[\sum_{\alpha=1}^N r_\alpha^2-\frac{1}{N}
\left(\sum_{\alpha=1}^N r_\alpha\right)^2\right]\right\}\cr
&&\exp\left\{-\frac{e^2 L|v-u|}{2}\left[l^2\left(\frac{N-1}{N}\right)+
\frac{2l}{N}\sum_{\alpha=1}^N
r_\alpha-2lr_\beta\right]\right\}\ \ .
\label{cortorov}
\end{eqnarray}
Eqs.(\ref{cortorov}) is different from the corresponding one obtained in
Ref.\cite{wipf}. As a matter of fact the restriction on the
integers $r_\alpha$ introduced by
the determinant factors, not only changes the spectrum of the theory but
also alters the asymptotic behavior of the correlators. 
Consider for example the case of the correlator of a quark-antiquark pair, $l=1$. As is known, any non-trivial theta-sector is unstable \cite{wit2,psz1,psz2} in this case since the quark-antiquark pair is energetically 
favoured to reduce the associated background electric field. 
The asymptotic behavior of Eq.(\ref{cortorov}) can be obtained by
sending $\tau\to\infty$. In this limit only the lowest representation
survives in the sums in Eq.(\ref{cortorov}).
The lowest
value for the quantity $\sum_\alpha(r_\alpha)^2-(\sum_\alpha
r_\alpha)^2/N$ can be obtained by choosing $r_\alpha=-\alpha$
(which corresponds to the singlet representation) and is given by the
curvature of the group manifold $N(N^2-1)/12$ as in Eq.(\ref{casi}).
The large $\tau$ limit of Eq.(\ref{cortorov}) is
\begin{eqnarray}
&&\lim_{\tau\to\infty}P_{1,-1}(u,v)=\left(\sum_{\beta=1}^N
-\sum_{\beta,\gamma}\delta_{\beta,\gamma+1}\right)
\exp\left[-\frac{e^2 L|v-u|}{2}\left(2\beta -\frac{1}{N}-N\right)\right]
=\cr&&=
\exp\left(-\frac{e^2 L|v-u|}{2}\frac{N^2-1}{N}\right)
\end{eqnarray}
We then have in this sector a confining potential between the
quark-antiquark pair with string tension $e^2(N^2-1)/N$. 

\subsection{Results for $SU(2)$}

Let us now study in particular the $SU(2)$ case on the torus.
For $N=2$ Eq.(\ref{cortorov}) reads
\begin{eqnarray}
P_{l,-l}(\theta_k;u,v)=  \frac{2\sqrt{2\pi}}{Z_k}\left\{
\sum_{\{r\ne-2\}=-\infty}^{+\infty}
\sum_{n=0}^{1}
e^{-\pi i n [k-1-r]}\right.\cr\left.
\exp\left\{-\frac{e^2 L\tau}{4}\left[(r+2)^2+l^2|v-u|\right]\right\}
\cosh\left[\frac{e^2 L(v-u)l(r+2)}{2}\right]\right.\cr\left.
-\sum_{n=0}^{1}
e^{-\pi i n (k-1-l)}
\exp\left\{-\frac{e^2L\tau l^2}{4}\left[1
-\frac{|v-u|}{\tau}\right]\right\}\right\}\ \ ,
\label{cortosu2}
\end{eqnarray}
Recall that the integers $r+1$ for $r\ge-1$ or $-r-3$ for $r\le-3$, in
Eq.(\ref{cortosu2}) give the 
number of boxes in the rows of the Young table in a given
representation. Consequently, for $k=0$ only representations whose Young
tableaux have an even
number of boxes are present in the sum, for $k=1$ only those with an odd
number.
Identifying $j=(r+1)/2$ for $r\ge-1$ and $j=-(r+3)/2$ for $r\le-3$, 
$j=0,1/2,1\dots$ we can
rewrite the loop
correlator for $SU(2)$ as
\begin{eqnarray}
&&P_{l,-l}(k=0;u,v)=  \frac{2\sqrt{2\pi}}{Z_0}\left\{4
\sum_{\{j=0,1,\dots\}}^{+\infty}
\exp\left\{-{e^2 L\tau}\left[j(j+1)+\frac{1}{4}
+\frac{l^2|v-u|}{4\tau}\right]\right\}\right.\cr&&\left.
\cosh\left[\frac{e^2 L(v-u)l(2j+1)}{2}\right]
-\sum_{n=0}^{1}
e^{\pi i n (1+l)}\exp\left\{-\frac{e^2L\tau l^2}{4}\left[1
-\frac{|v-u|}{\tau}\right]\right\}\right\}\ \ ,
\label{cortosu20}
\end{eqnarray}
and
\begin{eqnarray}
&&P_{l,-l}(k=1;u,v)=  \frac{2\sqrt{2\pi}}{Z_1}\left\{4
\sum_{\{j=1/2,3/2,\dots\}}^{+\infty}
\exp\left\{-{e^2L\tau}\left[j(j+1)+\frac{1}{4}
+\frac{l^2|v-u|}{4\tau}\right]\right\}\right.\cr&&\left.
\cosh\left[\frac{e^2 L(v-u)l(2j+1)}{2}\right]
-\sum_{n=0}^{1}
e^{\pi i n l}\exp\left\{-\frac{e^2L \tau l^2}{4}\left[1
-\frac{|v-u|}{\tau}\right]\right\}\right\}\ \ ,
\label{cortosu21}
\end{eqnarray}
Taking $\tau\to\infty$ in the fundamental representation with $l=1$,  
we can see that the non trivial theta
sector has an instability which creates a repulsive potential between the external charges. Moreover, the prescence of fundamental charges
results breaks the center symmetry that gives rise to the theta-sectors
in the first place.  For example, with a pair of fundamental charges,
one region of the torus carries fluxes with even numbers of boxes
while another carries fluxes with odd numbers of boxes.  In general 
one should sum over all theta-sectors in this case to obtain correct 
results for correlators. Turning to asymptotic behaviour,
using Eq.(\ref{ztorus}) it is easy to see that the $\theta=0$ sector 
produces the leading behaviour with inter-quark potential given by
\begin{equation}
\lim_{\tau\to\infty} -\frac{1}{L} \log
P_{1,-1}(0;u,v)=\frac{3e^2}{4}|v-u|\ \ .
\end{equation}

Let us consider now the correlator for a pair of adjoint $SU(2)$
loops. A loop in the adjoint representation can be taken
as the modulus squared trace of the fundamental representation group
element
\begin{equation}
{\rm Tr} U_{\rm adj}(u)=\left|{\rm Tr}U(u)\right|^2 - 1\ \ .
\end{equation}
For $SU(2)$,
\begin{equation}
{\rm Tr} U_{\rm adj}(u)={\rm Tr}U^2(u)+1\ \ .
\end{equation}
For the correlator we then have
\begin{equation}
\left<{\rm Tr} U_{\rm adj}(u){\rm Tr} U_{\rm adj}(v)\right>=1+
\left<{\rm Tr}U^2(u) {\rm Tr} U^2(v)\right>+
\left<{\rm Tr}U^2(u)\right> +\left<{\rm Tr} U^2(v)\right>\ .
\end{equation}
Using Eqs.(\ref{cortoro}) and (\ref{cortoro1})
\begin{eqnarray}
&&\left<{\rm Tr} U_{\rm adj}(u){\rm Tr} U_{\rm adj}(v)\right>_k=1 - 8
\frac{\sqrt{2\pi}}{Z_k}\sum_{n=0}^1\exp\left[-\frac{e^2 L\tau}{4}-i \pi n
k\right]\cr
&&+4\frac{\sqrt{2\pi}}{Z_k}\left\{
\sum_{n=0}^1\sum_{r=0}^\infty\exp\left[-\frac{e^2
L\tau (r+1)^2}{4}-i \pi n
(k-r)\right]\left\{\exp\left[{e^2 L|v-u|r}\right]
\right.\right.\cr&&\left.\left.
+\exp\left[-{e^2 L|v-u|(r+2)}\right]\right\}
-\sum_{n=0}^1\exp\left[-{e^2 L\tau}+{e^2 L |v-u|}-i \pi n
(k-1)\right]\right\}
\label{adjcor}
\end{eqnarray}
As before, only representations with an even number of boxes,
$r$ even, will contribute when $k=0$ and representations with an odd
number of boxes
when $k=1$. 
Taking into account the behavior of $Z_k$, Eq.(\ref{ztorus}), for large
$\tau$, we obtain a confining behavior in the trivial theta-sector
\begin{equation}
\lim_{\tau\to\infty} -\frac{1}{L} \log
\left<{\rm Tr} U_{\rm adj}(u){\rm Tr} U_{\rm adj}(v)\right>_0=
2 e^2 |v-u|\ \ ,
\end{equation}
and a screening behavior in the non-trivial theta-sector.
\begin{equation}
\lim_{\tau\to\infty}-\frac{1}{L}\log
\left<{\rm Tr} U_{\rm adj}(u){\rm Tr} U_{\rm adj}(v)\right>_1=
\frac{1}{L}\left[
1-\exp\left(-{3}{e^2 L}|v-u|\right)\right]\ \ .
\end{equation}
These results are in agreement with the general discussion of
ref.\cite{wit2}.

As a check of previous calculations, we will now 
evaluate these correlators for the case of 
$SU(2)$, using the character representation.  
We start by considering the general case of the correlator
of a pair of Polyakov loops in (irreducible) representations
$R$ and $R^\prime$  
on a cylinder
\beq
P_{R,R^\prime}(U_1, U_2; u,v) = \frac{1}{Z} 
\int [dV_1] [dV_2] 
K[u; U_1, V_1] \chi_R(V_1) K[v-u; V_1, V_2] 
\chi_{R^\prime}(V_2) K[ \tau-v; V_2, U_2]  
\eeq
Using the definition of the kernel $K$ (\ref{hkchar})
and the properties of 
the group characters this expression can be reduced to a sum 
over representations for $v>u$
\bea
P_{R,R^\prime}(U_1, U_2; u,v)& =& \label{cylchar}\\
\frac{1}{Z}  \sum_{R_1, R_2, R_3}&  & \!\!\!\!\!\!\!\!\!\!\!\!\!
N_{R R_2}^{R_1} N_{R_3 R^\prime}^{R_2}
\chi_{R_1}(U_1) \chi_{R_3}(U_2) 
\e^{-\{e^2 L [u C_2(R_1)+(v-u) C_2(R_2)+(\tau-v) C_2(R_3)]\}}
\nonumber
\eea
Using this result
(\ref{cylchar}) we can 
easily find the analogous on the torus.
We can set $U_1=U_2=U$ and integrate on $U$ to 
obtain the  pair correlator of Polyakov loops on the torus
Using the properties of characters this reduces to 
\beq
P_{R,R^\prime}(u,v) = \frac{1}{Z}
\sum_{R_1 R_2} N^{R_1}_{R_2 R} N^{R_2}_{R_1 R^{\prime} }
\exp{ \left( -{e^2 L |v-u|} C_2(R_1) -{e^2 L (\tau}-|u-v|) C_2(R_2) \right) } 
\eeq

From this general formula one can immediately make quantitative
statements about the binding between pairs of loops in the case where
one side of the torus becomes large $(\tau \rightarrow \infty)$. In
the trivial theta-sector (recall this means we consider the sum of all
distinct theta-sectors)
\beq
P_{R,R^\prime}(u,v) \rightarrow \delta_{R R^\prime}
\e^{-e^2 L |v-u| C_2(R)}
\label{trivial}
\eeq

As well one can easily do the same for the correlator of a pair of
Polyakov loops in the $k^{\rm th}$ theta-sector
\bea
P_{R,R^\prime}(\theta_k; u,v)  &&= \\
\frac{1}{Z_k} \sum_{R_1 R_2}  
& &\!\!\!\!\!\!N^{R_1}_{R_2 R} N^{R_2}_{R_1 R^{\prime} } 
\delta_N(k, C_1(R_1) )
\exp{ \left( -{e^2 L|v-u|} C_2(R_1) -
e^2 L(\tau-|v-u|) C_2(R_2) \right) } 
\nonumber
\eea
In the case of two Polyakov loops in the adjoint representation some
calculation \cite{psz1} shows for $N-1 >k>1$, the pair correlator is
\beq
P_{\rm{Ad},\rm{Ad}}(\theta_k; u,v)  \rightarrow (1 + 
\e^{-e^2 L|v-u| k}+\e^{-e^2 L|v-u| (N-k)} +\e^{-e^2 L |v-u| (N+1)} )
\eeq
The cases $k=1$ and $k=N-1$ are given by excluding the second and
third terms respectively. The case $k=0$ recovers the topologically
trivial case (\ref{trivial}) in this limit.

Let us now consider in particular the $SU(2)$ case where the fusion
numbers are known explicitly.  We will label representations by the
single non-negative integer $n$ which is equal to $l_1$ in terms of
row variables or $m_1$ in terms of column variables in the associated
Young table.  Consequently, $C_2(n) = ((n+1)^2 -1)/4$.  The fusion
numbers are
\begin{equation}
N^i_{jl}=\left\{\matrix{  1&{\rm when}~  
i=j+l,j+l-2,\ldots, |j-l| \cr
0&{\rm otherwise}\cr} \right.
\end{equation}
 
Bringing these facts together we have the correlator of pair of
Polyakov loops, one in representation $n$ and the other in $m$,
separated by distance a $|v-u|$
\bea
P_{n,m}(u,v) &=&  \label{gencorr}\\
\!\!\!\!\!\!\!\!\!\!\frac{1}{Z}
\sum^{\infty}_{j,l=0} \!\!\!\!&&\!\!\!\!\!\!\!\!\!\!\!
\e^{- \frac{e^2 L|v-u|}{4} [(l+1)^2 -1]}
\e^{- \frac{e^2 L\tau}{4} [(j+1)^2 -1]} 
\sum^{\min{l,m}}_{r=0} \sum^{\min{j,s}}_{s=0} 
\delta_{m+n, 2(s+r)} \delta_{2l +2s +m, 2j + 2 r+n}
\nonumber
\eea
The first delta function serves to enforce the condition that 
$m+n$ must be even. This is a particular example of the general 
fact in $SU(N)$ that in order for the 
correlator of any system of loops to be non-vanishing the total charge
of the loops must be vanishing mod $N$.  This restriction ensures
that the system contains a charge singlet.   
 
Two special cases that are of interest are the pair correlators of 
fundamental and adjoint Polyakov loops. For the fundamental case
$n=m=1$ and
\beq
P_{1,1}(u,v) = \frac{1}{Z} 
\sum^{\infty}_{l=0} 
\e^{- \frac{e^2 L\tau}{4}[(l+1)^2 -1]} [ 
\e^{- \frac{e^2 L|v-u|}{4} (2l+3)} +  
\e^{-\frac{e^2 L(\tau-|v-u|)}{4} (2 l+3) }]
\eeq
It can be checked that this result coincides with the sum of Eqs.
(\ref{cortosu20}) (\ref{cortosu21})
 when the different zero point energy and normalization of $Z$
in the two approaches, are taken into account.
Likewise for the correlator of a pair of adjoint loops $m=n=2$
\bea
P_{2,2}(u,v) &=&  \frac{1}{Z} \left(
\e^{- {2 e^2 L|v-u|}}  + \e^{-2 e^2 L(\tau-|v-u|)}
\right. \\
&& \left. +
\sum^{\infty}_{l=1} 
\e^{- \frac{e^2 L\tau}{4} [(l+1)^2 -1]} [ 
\e^{- {e^2 L|v-u|} ( l+2)} +  
1 + \e^{-e^2 L(\tau-|v-u|) (l+ 2) }] \right)
\nonumber
\eea

Now we turn the issue of non-trivial $\theta$-sectors in the 
correlators.  As usual this is carried out by including a projection 
operator.  For $SU(2)$ this prescription amounts to restricting the
sum over representations to those with either odd or even numbers
of boxes in the corresponding Young table. It should be noted that
this can be done in a consistent manner only if
each of the charges in the system has vanishing 2-ality 
(ie. has $2n$ boxes).
 
For the correlator of a pair of adjoint 
Polyakov loops the two cases are $k=0$ where one
would sum over even representations in (\ref{gencorr}) 
\bea
P_{2,2}(\theta_0;u,v) &=&  \frac{1}{Z_0} \left(
\e^{-2 e^2 L|v-u|}  + \e^{-2 e^2 L(\tau-|v-u|)} \right. \\
&& \left. + 
\sum^{\infty}_{l=2,4,\cdots} 
\e^{- \frac{e^2 L\tau}{4} [(l+1)^2 -1]} [ 
\e^{- {e^2 L|v-u|} ( l +2)} +  
1 + \e^{-e^2 L(\tau-|v-u|) (l + 2) }] \right)
\eea
and  the case $k=1$ where the sum is over odd representations
\beq
P_{2,2}(\theta_1;u,v) = \frac{1}{Z_1}  
\sum^{\infty}_{l=1,3,\cdots} 
\e^{- \frac{e^2 L\tau}{4} [(l+1)^2 -1]} [ 
\e^{- {e^2 L|u-v|} (l+2)} +  
1 + \e^{-e^2 L(\tau-|v-u|) (l + 2) }] 
\eeq
These results coincide with the results of the functional integral technique, Eq.(\ref{adjcor}).

\section{Discussion}

We have demonstrated the utility of the diagonalization method for
doing practical computations in 2D Yang-Mills theory.  This method is
alternative to the well-known character expansion technique.  The
diagonalization method is useful for computing correlators of higher
winding loop operators.  Knowing the formulae for higher
representations in terms of higher winding loops (via characters), and
the formulae for correlators in the character representation, it
should be possible to obtain information about the fusion rules for
higher representations of SU(N) groups (and in fact, by suitable
generalization of our work, arbitrary Lie groups) from our formulae.
We have not pursued this problem in the present work.

\noindent
\section* {Acknowledgments} G. Grignani and P. Sodano acknowledge the hospitality
of the Physics Department of the University of British Columbia where
this work was completed.  G. Semenoff thanks the INFN and the Universita
di Perugia for hospitality which enabled the beginning of this collaboration.


\begin{thebibliography}{}

\bibitem{wit} { \sc E. Witten}, {\sl Commun. Math. Phys.} {\bf 141} (1991), 153.
 
\bibitem{blau0} {\sc M. Blau and G. Thompson}, {\sl Proceedings 
of Summer School in High Energy Physics and Cosmology, June 1993,
Trieste HEP Cosmol.} (1993), 175.
 
\bibitem{blau} {\sc M. Blau and G. Thompson}, {\sl Int. J. Mod. Phys.}
{\bf A7} (1992), 3781.

\bibitem{cmr} {\sc S.Cordes, G.Moore and S.Ramgoolam}, {\sl Proceedings of Les Houches Summer School and Trieste Spring String School, 1994},
hep-th/9411210.

\bibitem{ab}{\sc E. Abdalla and M.C.B. Abdalla}, {\sl Phys.Rept.} {\bf 265} (1996), 253.

\bibitem{mig} {\sc A. A. Migdal}, {\sl Sov. Phys. JETP} {\bf 42} 
(1975), 413.  

\bibitem{rus} {\sc B. Y. Rusakov}, {\sl Mod. Phys. Lett.} {\bf A5}
(1990), 693.

\bibitem{gt1} {\sc D.J.Gross and W. Taylor}, {\sl Nucl. Phys.} {\bf B400} (1993), 191.

\bibitem{gt2} {\sc D.J.Gross and W. Taylor}, {\sl Nucl. Phys.} {\bf B403} (1993), 395.

\bibitem{mp1} {\sc J.Minahan and A. Polychronakos}, {\sl Phys. Lett.} {\bf B212} (1993), 155.

\bibitem{mp2} {\sc J.Minahan and A. Polychronakos}, {\sl Nucl. Phys.} {\bf B422} (1994), 172.

\bibitem{dadda} {\sc M. Caselle, A. D'Adda, L. Magnea and S. Panzeri}, {\sl Nucl. Phys.} {\bf B416} (1994), 751.

\bibitem{hh1} {\sc J. E. Hetrick}, {\sl Nucl. Phys. Proc. Suppl.} 
{\bf B34} (1994), 805.

\bibitem{hh2} {\sc J. E. Hetrick}, {\sl Int. J. Mod. Phys.} 
{\bf A9} (1994), 3153.

\bibitem{hh3} {\sc J. E. Hetrick}, {\sl Nucl. Phys. Proc. Suppl.} 
{\bf B30} (1993), 228.

\bibitem{hh4} {\sc J. E. Hetrick and Y. Hosotani}, {\sl Phys. Lett.} {\bf B230} (1989), 88.

\bibitem{gss} {\sc G. Grignani, G. Semenoff and P. Sodano}, hep-th/9503109.

\bibitem{wipf} {\sc U. G. Mitreuter, J. M. Pawloski and A. Wipf},
hep-th/9611105.

\bibitem{r3} {\sc S. G. Rajeev}, {\sl Phys. Lett.} {\bf B217} (1989),
123.

\bibitem{r1} {\sc S.G. Rajeev and L. Rossi}, {\sl J. Math. Phys.} 
{\bf 36} (1995), 3308.

\bibitem{r2} {\sc K. S. Gupta, R. J. Herdersson, S. G. Rajeev and O. T. Turgot},
{\sl J. Math. Phys.} {\bf 35} (1994), 3845.

\bibitem{gsst1} {\sc G. Grignani, G. Semenoff, P. Sodano and O. Tirkkonen}, {\sl Nucl. Phys.} {\bf B489} (1997), 360.

\bibitem{gsst2} {\sc G. Grignani, G. Semenoff, P. Sodano and O. Tirkkonen}, {\sl Nucl. Phys.} {\bf B473} (1996), 143.

\bibitem{gsst3} {\sc G. Grignani, G. Semenoff, P. Sodano and O. Tirkkonen}, {\sl Int. J. Mod.
Phys.} {\bf A11} (1996), 4103.
 
\bibitem{gps} {\sc C. Gattringer, L. Paniak and G. Semenoff}, {\sl Ann. Phys.}  In press, hep-th/9612030.

\bibitem{pol} { \sc A. M. Polyakov}, {\sl Phys. Lett.} {\bf 72B} (1978), 477.

\bibitem{sus} { \sc L. Susskind}, {\sl Phys. Rev.} {\bf D20} (1979), 2610.

\bibitem{sy1} { \sc B. Svetitsky and L. Yaffe}, {\sl Nucl. Phys.} 
{\bf B210} (1982), 423.

\bibitem{sy2} { \sc B. Svetitsky}, {\sl Phys. Rept.} {\bf 132} (1986), 1.

\bibitem{blau2} {\sc M. Blau and G. Thompson}, {\sl J. Math. Phys.} {\bf 36} (1995), 2192. 

\bibitem{stz} { \sc G. W. Semenoff, O. Tirkkonen and K. Zarembo},
{\sl Phys. Rev. Lett.} {\bf 77} (1996), 2174. 

\bibitem{wit2}{\sc E. Witten}, {\sl Il Nuovo Cimento} {\bf 51A} (1979), 325.
y
\bibitem{psz1} {\sc L. Paniak, G. Semenoff and A. Zhitnitsky}, 
{\sl Nucl. Phys.} {\bf B487} (1997), 191.

\bibitem{psz2} {\sc L. Paniak, G. Semenoff and A. Zhitnitsky}, hep-ph/9701270.

\bibitem{coleman1} {\sc S. Coleman, R. Jackiw and L. Susskind}, {\sl 
Ann. Phys.} {\bf 93} (1975), 267.

\bibitem{coleman2} {\sc S. Coleman}, {\sl Ann. Phys.} {\bf 101} (1976),
239.

\bibitem{ls1} { \sc E. Langmann and G. Semenoff}, {\sl Phys. Lett.} {\bf B296}
(1992), 117.

\bibitem{ls2} { \sc E. Langmann and G. Semenoff}, 
{\sl Phys. Lett.} {\bf B303} (1993), 303.

\bibitem{kmsw}{\sc I. I. Kogan, A. Morozov, G. Semenoff and  N. Weiss}, {\sl Nucl. Phys.} {\bf B395} (1993), 547.

\end{thebibliography}
\end{document}